\documentclass[preprint,12pt]{elsarticle}

\usepackage{amssymb}
\usepackage{amsmath}
\usepackage[table]{xcolor}
\usepackage{array}
\usepackage{multicol}
\usepackage{setspace}
\usepackage[normalem]{ulem}
\usepackage{hyperref}

\usepackage{graphicx}
\usepackage{multirow}
\usepackage{color}
\usepackage{url} 

\journal{Journal of Systems and Software}

\begin{document}

\begin{frontmatter}



\title{Exploring Developer Experience Factors in Software Ecosystems}


\author[1,2]{Rodrigo Oliveira Zacarias} 
\ead{rodrigo.zacarias@edu.unirio.br}

\author[1]{Léo Carvalho Ramos Antunes} 
\ead{leo.antunes@edu.unirio.br}

\author[1]{Márcio de Oliveira Barros} 
\ead{marcio.barros@uniriotec.br}

\author[1]{Rodrigo Pereira dos Santos} 
\ead{rps@uniriotec.br}

\author[3]{Patricia Lago} 
\ead{p.lago@vu.nl}

\affiliation[1]{organization={Federal University of the State of Rio de Janeiro (UNIRIO)},
            city={Rio de Janeiro},
            country={Brazil}}

\affiliation[2]{organization={Fluminense Federal University (UFF)},
            city={Niterói},
            country={Brazil}}

\affiliation[3]{organization={Vrije Universiteit Amsterdam},
            city={Amsterdam},
            country={The Netherlands}}

\begin{abstract}

\textbf{Context:} Developer experience (DX) plays a key role in developers' performance and their continued involvement in a software ecosystem (SECO) platform. While researchers and practitioners have recognized several factors affecting DX in SECO platforms, a clear roadmap of the most influential factors is still missing. This is particularly important given the direct impact on developers’ interest in SECO and their ongoing engagement with the common technological platform.
\textbf{Goal:} This work aims to identify key DX factors and understand how they influence third-party developers' decisions to adopt and keep contributing to a SECO.
\textbf{Methods:} We conducted a systematic mapping study (SMS), analyzing 29 studies to assess the state-of-the-art of DX in SECO. Additionally, we conducted a Delphi study to evaluate the influence of 27 DX factors (identified in our SMS) from the perspective of 21 third-party developers to adopt and keep contributing to a SECO.
\textbf{Results:} The factors that most strongly influence developers' adoption and ongoing contributions to a SECO are: ``financial costs for using the platform'', ``desired technical resources for development'', ``low barriers to entry into the applications market'', and ``more financial gains''.
\textbf{Conclusion:} DX is essential for the success and sustainability of SECO. Our set of DX factors provides valuable insights and recommendations for researchers and practitioners to address key DX concerns from the perspective of third-party developers.
\end{abstract}
%
%




\end{frontmatter}



\section{Introduction}\label{sec:introduction}

In a scenario of investments in software development strategies and approaches to meet new market demands, it has been a great challenge for corporations to maintain a system/software architecture fully internalized to its organization~\cite{Barbosa2013}. For this reason, some companies have invested in opening their architectures to allow third-party developers to collaborate in producing their components over a common technological platform. That practice defines the notion of a software ecosystem (SECO)~\cite{Jansen2009,santos2016}.

Bosch~\cite{Bosch2009} states that two reasons motivate an organization to use the SECO approach: (i) it may realize that the amount of functionality required to develop to satisfy customers' and users' needs is much greater than it can build; and (ii) the trend towards mass customization drives demand for investment in research and development for successful software applications, so that extending the product (and the platform) with support of external actors to the organization, mainly third-party developers, seems to be an effective mechanism.

Whether in the more traditional approaches to software development or approaches based on the concept of SECO developing and creating software is an activity that requires both technical and social skills from developers. Software developers are creators and designers when they write the code and design the logic that structures the software, but they are also users of the tools they use in their craft. It will result in a type of user experience (UX), giving rise to the concept of developer experience (DX)~\cite{Nylund2020}. 

Fagerholm and Münch~\cite{Fagerholm2012} define DX as a broad concept that captures how developers feel about, think about, and value their work. In other words, DX is a term that explains how developers interact with the developing software practice, both technically and socially~\cite{Nylund2020}. In the context of SECO, DX is a decisive coefficient both for the performance of developers and for keeping developers actively contributing to SECO's technological platform \cite{FontaoEtAl2017}.

Maintaining third-party developers' engagement is one of the decisive factors for SECO health, i.e., the ability of an ecosystem to maintain productivity, robustness, and niche creation over time~\cite{Simone2017}. In many cases, engagement is related to DX. Fun experience, intellectual stimulation, and learning new skills are some of the main motivations for developers to interact within a SECO~\cite{Koch2014, Fontao2020}. An unsatisfactory DX in the SECO environment can lead to a lack of interest and engagement among third-party developers to the platform. More severe cases can culminate in the SECO's ``death''~\cite{Fontao2020}.

Despite the importance of the DX concept for software development, the definition of such a concept is still incipient. In many types of research in academia and industry, the content and topic of the results vary much~\cite{Nylund2020}. While researchers and practitioners have recognized several factors affecting DX in SECO platforms, a clear roadmap of the most influential factors is still missing. This is particularly important given the direct impact on developers’ interest in SECO and their ongoing engagement with the common technological platform~\cite{STEGLICH2023111808, Parracho2023, fontao2021developer}. Concerning this, we can formulate the following research question (RQ): \textit{``How do DX factors influence third-party developers to adopt and keep contributing to a SECO?''} 

This work aims to identify key DX factors and understand how they influence third-party developers' decisions to adopt and keep contributing to a SECO. To do so, we conducted a systematic mapping study (SMS), analyzing 29 studies to assess the state-of-the-art of DX in SECO. Additionally, we conducted a Delphi study to evaluate the influence of 27 DX factors (identified in our SMS) from the perspective of 21 third-party developers to adopt and keep contributing to a SECO. As the main contribution, our list of DX factors, organized into four categories (Common Technological Platform, Projects and Applications, Community Interaction, and Expectations and Value of Contribution), provides valuable insights and recommendations for researchers and practitioners to address key DX concerns from the perspective of software developers.

Regarding implications, researchers can find in this work an overview of what has been studied so far about DX related to the SECO context. The categorization of factors can function as a starting point for future research on more specific factors, such as exploring only factors related to development infrastructure or ecosystems. By identifying the factors, practitioners can improve DX by knowing what affects their daily lives and understanding the effects in SECO. Our set of DX factors can be used as a reference guide and recommendations to support practitioners from organizations dealing with the key concerns related to DX in SECO.

The remainder of this article is organized as follows: Section~\ref{sec:background} presents some background information, key concepts, and related work. Section~\ref{sec:method} describes the research method. Section~\ref{sec:results} shows the results of the studies. Furthermore, the discussion and implications of the results of both studies are presented in Section~\ref{sec:discussion}. Finally, Section \ref{sec:threats} presents the threats to this work's validity, and Section~\ref{sec:finalRemarks} highlights the conclusion and future work directions.

\section{Background}\label{sec:background}
This section describes the concepts related to SECO and DX. In addition, we also present related work to this research.

\subsection{Software Ecosystems}\label{sec:seco}
A SECO can be defined as a set of actors and their relationships that function as a unit, engaging with a distributed market of software and services. These relationships are often supported by a technological platform or a common market and are realized through the exchange of information, resources, and artifacts~\cite{Jansen2009}. To facilitate these exchanges, it is essential to integrate support mechanisms and tools to ensure communication and interaction between developers and users~\cite{santos2016, JANSEN2020}.

Considering the different relationships that those actors can have with a platform, three key roles are identified by Hanssen~\cite{Hanssen2012}: (i) keystone: an organization or group that leads the development of the platform; (ii) end-users: represents those who need the platform to run their own business; and (iii) third-party developers: use the platform as a basis to produce related solutions or services. The platform provider is not always present in the SECO as a single organization and can be represented as an open-source software community~\cite{Jansen2013}.

According to Manikas~\cite{Manikas2016}, SECO can be classified into three types: proprietary, open-source, and hybrid. Proprietary SECO (PSECO) have their value creation based on proprietary contributions confidentiality agreements protect the source code and other artifacts produced (e.g., SAP - System Applications and Products\footnote{https://www.sap.com/} - and AWS - Amazon Web Services\footnote{https://aws.amazon.com/})~\cite{Costa2022}. Open-source SECO (OSSECO) allow contributions from different actors and communities and actors generally do not participate to obtain direct income from their activity in the ecosystem (e.g., Eclipse Foundation\footnote{https://www.eclipse.org/org/foundation/}, GitHub\footnote{https://github.com/}, GitLab\footnote{https://about.gitlab.com/}, and Apache Foundation\footnote{https://www.apache.org/})~\cite{FRANCOBEDOYA2017160}. Finally, hybrid SECO support both proprietary and open-source contributions, such as using proprietary strategies to drive policy on the technological platform and using open-source strategies for community engagement, such as tools, submissions, and publishing contributions (e.g., Android\footnote{https://developer.android.com/} and iOS\footnote{https://developer.apple.com/})~\cite{Fontao2015}.

\subsection{Developer Experience}\label{sec:dx}
The definition of DX is influenced by the UX concept. ISO standard 9241-210:2019~\cite{ISO9241-210} defines UX as ``a person's perceptions and responses that result from the use or anticipated use of a product, system or service''. Although DX and UX may seem similar at first glance, DX is not just ``UX for developers''. Instead, DX is an extension of UX focused on users who build with technical languages and tooling. DX follows the same core principles of UX but extends it by recognizing that technical details and mechanical processes can be understood and utilized efficiently by a developer. The distinction between UX and DX comes from the different needs that end-users and developers have \cite{Fagerholm2012, Greiler2022}.

Greiler et al.~\cite{Greiler2022} define DX as ``how developers think about, feel about, and value their work''. Their definition is inspired by Fagerholm and Münch \cite{Fagerholm2012}, who define DX in terms of the theory of the trilogy of mind from social psychology~\cite{Hilgard1980}. The three main dimensions of the mind are cognition, emotion, and expectation (also referred to as conation). Based on this theory, Fagerholm and Münch~\cite{Fagerholm2012} elaborated a conceptual framework to define DX.

According to the framework, the DX concept is composed of the following dimensions: (i) cognitive dimension: consists of factors that affect how developers perceive their development infrastructure at an intellectual level (platform, techniques, process, skills, and procedures); (ii) affective dimension: consists of factors that influence the way developers feel about their work (respect, social, team, attachment, and belonging); and (iii) conative dimension: consists of factors that affect how developers see the value of their contributions (plans, goal, alignment, intention, motivation, and commitment).

The evaluation of DX is important for estimating the factors that can impact the productivity of development teams, as well as understanding how knowledge is applied and shared among members. In the context of SECO, understanding these factors helps optimize developer engagement and collaboration, especially in attracting new third-party developers to a platform.

Fontão et al.~\cite{FontaoEtAl2017} indicate that DX plays a significant role in the productivity of developers when publishing an application in a SECO. Additionally, attracting and engaging third-party developers to a common technological platform is necessary for the continuity and growth of SECO. To achieve this, it is important to understand the expectations, perceptions, and feelings of these developers, particularly during the onboarding and engagement process, to ensure efficient and sustainable collaboration with the ecosystem.

Although the concept of DX was introduced by Fagerholm and Münch~\cite{Fagerholm2012}, there is still room for investigating the specific factors that influence DX in SECO. In a SECO, third-party developers play a crucial role in addressing the internal limitations of platforms, contributing to the maintenance and evolution of the SECO's strategy, as well as to the improvement of the SECO's health indicators~\cite{fontao2021developer}. 

Furthermore, DX is influenced by platform quality attributes, such as transparency. Transparency refers to a set of characteristics on the visibility and clarity of information within the organizational or collaborative environment, allowing all stakeholders to understand how decisions are made and how processes function. To keep developers active and engaged, it is also important to investigate the relationship between transparency characteristics and DX. Transparency solutions designed to improve DX in SECO can help increase the attractiveness and engagement of new third-party developers, which is important for the expansion and long-term sustainability of a SECO platform~\cite{ZacariasEtAl2024}.

\subsection{Related Work}\label{sec:relatedWork}
In our searches, we have identified some studies in the literature that explored factors that affect developers in SECO and/or software development. Greiler et al.~\cite{Greiler2022} explored factors that affect DX in development teams. To do so, the authors reviewed the literature and then conducted semi-structured interviews with developers from the industry, which they transcribed and iteratively coded. Their findings highlighted factors that affect DX and characteristics that influence their respective importance to individual developers. As the main contribution, the authors developed an actionable framework to organize those factors and, in addition, present strategies employed by individuals and teams to improve DX, the barriers that stand in their way, and the coping mechanisms of developers when DX cannot be sufficiently improved. 

Greiler et al.~\cite{Greiler2022} present a solid overview of DX in software development teams. However, when it comes to SECO, other characteristics need to be considered to understand DX in such a context. In SECO, a keystone does not have full control over the developers contributing to a common technological platform. This situation is a significant challenge because developers operate more independently, which can make it difficult to align their interests with the keystone’s goals. Rather than direct control, a keystone needs to promote collaboration through incentives and norms, creating a balance between autonomy and strategic direction within its ecosystem. This control varies considering the different types of SECO. In an OSSECO, a keystone has little control, allowing developers to join and leave its common technological platform freely. In a PSECO, a keystone exerts more control over who can contribute to its platform, setting strict rules. In a hybrid SECO, there is a balance between control and flexibility, since a keystone defines guidelines but also allows third-party participation. Therefore, this dynamic characteristic of SECO and these reasons need to be explored further to characterize DX in SECO.

Steglich et al.~\cite{STEGLICH2023111808} investigated factors that affect developers’ decision to participate in a mobile software ecosystem (MSECO). First, the authors analyzed the literature to identify such motivational factors and conducted interviews to identify the opinions of MSECO developers regarding the list of 29 factors identified, classified into social, economic, and technical aspects. These factors are mainly related to aspects and characteristics of the SECO platform and to the developers' own expectations and motivations, such as ``To possess the desired technical resources'', ``To learn and improve skills'', ``To become recognized by the community'', among others. Their results indicated that the lack of studies focusing on developers is a key to understanding developers’ decision to participate in an MSECO. Moreover, developers become more concerned with their relationships in these ecosystems over time. This situation affects MSECO sustainability, which depends on two capabilities: a SECO to adapt itself, when necessary, to new technologies and resources; and to attract and retain developers who are interested in contributing to a SECO. Therefore, this list helps a keystone to understand which motivating factors can attract third-party developers and contribute to the sustainability of MSECO.

While Steglich et al.~\cite{STEGLICH2023111808} focused exclusively on analyzing the context of MSECO (a specific type of hybrid SECO), our study adopts a broader approach by investigating DX factors common across different types of SECO: proprietary, open-source, and hybrid. Additionally, although Steglich et al.~\cite{STEGLICH2023111808} classified the factors into social, economic, and technical categories, the authors did not analyze these factors from a DX perspective. In contrast, our work expands this analysis by using the DX framework proposed by Fagerholm and Münch~\cite{Fagerholm2012}, as described in Section~\ref{sec:dx}, which encompasses the categories used in that work and also considers developers’ expectations regarding a SECO platform.

Gonçalves et al.~\cite{Gonçalves2024} investigated the factors, barriers, and strategies influencing DX in software reuse. According to the authors, there is a gap in the literature and industrial practices regarding the real experience of developers in software reuse. Many approaches focus primarily on technical aspects, overlooking the perspective of developers who often encounter psychological barriers. In light of this issue, the authors selected 10 studies through a rapid review for detailed data extraction. Their findings identified 15 factors affecting DX in software reuse, 7 barriers that impede developers from improving DX, and 13 strategies to enhance it. Their results highlighted the critical role of comprehensive documentation, a clear understanding of software functionality, and robust reuse-compatible infrastructure as key technical factors. Organizational support, effective resource allocation, and fostering a communication, collaboration, and self-efficacy culture were essential for successful software reuse. As implications for researchers and practitioners, the authors offered practical guidance to develop more effective reuse practices and improve DX.

The context of software reuse investigated by Gonçalves et al.~\cite{Gonçalves2024} has similar characteristics to SECO. We can consider that SECO is the next stage of software reuse~\cite{Bosch2009}. However, the authors indicated that the relationship between the factors within the software industry still requires further investigation. In our work, in addition to reviewing the literature, we discussed DX factors with third-party software developers in a Delphi study to understand their perspective on the influence of the factors on developers’ interaction with SECO. As such, we also aim to fill the gap identified by Gonçalves et al.~\cite{Gonçalves2024}, going beyond the technical aspects of a SECO platform and considering the expectations of developers in industrial practice within this context.

Finally, in a previous work, we investigated and characterized conditioning factors for transparency in SECO~\cite{ZacariasEtAl2024}. To do so, we conducted an SMS with  23 selected studies and a field study with 16 software developers to identify and analyze such factors. During the analysis of the results, we noticed that transparency in SECO, in general, is a factor that impacts DX. For example, if a third-party developer wants to learn how to develop applications on a SECO but cannot easily access or understand it, such a developer may struggle to develop applications independently. This situation can lead to a negative DX and may cause such a developer to abandon their efforts to interact with the common technological platform.

Therefore, this type of situation has motivated us to understand more deeply which other factors can impact DX of these third-party developers. The focus on such third-party developers considers the fact that they are one of the key actors in SECO, responsible for expanding the common technological platform of these ecosystems. The expansion of a SECO is directly linked to its robustness and long-term sustainability in the software market.

\section{Research Method}\label{sec:method}
\begin{figure}[!ht]
    \centering
    \includegraphics[width=1.0\linewidth]{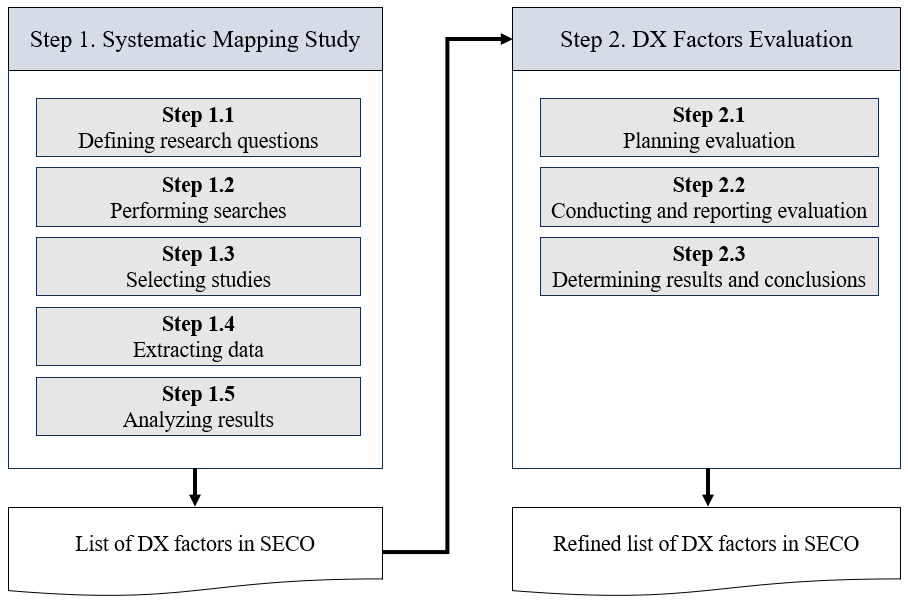}
    \caption{Our research method.}
    \label{fig:method}
\end{figure}

Following the guidelines of ACM SIGSOFT Empirical Standards~\citep{Ralph2021}, this research method is characterized as exploratory and follows quantitative and qualitative data collection and analysis approaches. This work aims to identify key DX factors and understand how they influence third-party developers' decisions to adopt and keep contributing to a SECO. To do so, we defined a research method consisting of two steps, as shown in 
Figure~\ref{fig:method}.

To answer our RQ \textit{``How do DX factors influence third-party developers to adopt and keep contributing to a SECO?''}, first, we conducted an SMS~\cite{Petersen2015} to identify a list of DX factors in SECO in the scientific literature. Then, we asked third-party developers to evaluate how that list of DX factors influences them to adopt and keep contributing to a SECO through a questionnaire in a Delphi study~\cite{DALKEY1969408}. The details of both studies and their respective sub-questions (SQ) are described in the following sections.

\subsection{Step 1: Systematic Mapping Study}\label{sec:sms}

In this step, we conducted a SMS to review the state-of-the-art of DX in SECO, following the guidelines proposed by Petersen et al~\cite{Petersen2015}. This protocol is structured in five steps: (1.1) Defining research questions; (1.2) Performing searches; (1.3) Selecting studies; (1.4) Extracting data; and (1.5) Analyzing results.

\subsubsection{Step 1.1: Defining research questions}\label{sec:defRQ}
This SMS aims to investigate factors that influence the DX in SECO. To achieve this goal, we proposed four SQ to help answer our RQ:

\begin{itemize}
\item {\bfseries{SQ1}}: Which factors affect DX in SECO?
\item {\bfseries{SQ2}}: Which research methods are used to obtain these factors?
\item {\bfseries{SQ3}}: Which types of ecosystems have been considered? 
\item {\bfseries{SQ4}}: Which concerns are pointed out for DX in SECO? 
\end{itemize}

The first sub-question (SQ1) aims to gather the most relevant factors that influence the DX in SECO. We consider factors as circumstances that affect the individual, something they cannot control or change alone. The research methods applied for finding these factors are pinpointed by SQ2 (e.g., field study, interview, survey etc.). Furthermore, SQ3 identifies the characteristics of SECO addressed in those studies, such as type of SECO, name, technology etc. Finally, SQ4 points out the DX concerns and how these aspects can influence SECO. Concerns refer to issues and challenges frequently highlighted in the literature or by software industry professionals regarding a particular environment or domain. Through concerns, researchers and practitioners can identify and plan research opportunities~\citep{motta2018challenges}.

\subsubsection{Step 1.2: Performing searches}\label{sec:search}
The framework PICO (Population, Intervention, Comparison, and Outcomes), suggested by Kitchenham and Charters~\cite{KitchenhamAndCharters2007}, was followed to identify keywords and formulate search strings from research questions.~\textbf{Population}: In our context, the population comprises studies on SECO. \textbf{Intervention}: The intervention is DX.~\textbf{Comparison}: There is no clear comparison in the context of this study.~\textbf{Outcomes}: The outcome is the factors that influence DX.

To create our search string, we joined the keywords that represented the population and the intervention in the framework PICO. The \emph{factors} keyword was omitted to avoid missing important studies that do not cite this keyword or its synonyms explicitly. As our research concerns an SMS, there was no specific comparison nor the need to limit the search space regarding outcomes, following the adopted search strategy by Villamizar et al.~\cite{Villamizar2021}. The keywords were then grouped into sets with their synonyms and considered to formulate the search string.

\begin{itemize}
\item {\bfseries{Set 1}}: Scoping the search for SECO: ``software ecosystem'' or ``SECO'';
\item {\bfseries{Set 2}}: Search terms directly related to developer experience: ``developer experience'', ``developer motivation'', ``developer relation'', ``developer satisfaction'', or ``programmer experience''. 
\end{itemize}

Those two sets were used together to form the base search string, which was run on the following databases: Scopus, Science Direct, IEEE Xplore, Web of Science, ACM Digital Library, and Springer Link. These databases have been selected based on the recommendations of Dyba et al.~\cite{dyba2007applying}. The search strings used for each database can be found in Table~\ref{tab:dbsearches}. Furthermore, Table~\ref{tab:dbnumstudies} presents the number of search results per database. We used the works of Steglich et al.~\cite{STEGLICH2023111808} and Parracho et al.~\cite{Parracho2023} as control studies to evaluate and refine our search string.

\begin{table}[!ht]
  \scriptsize
  \centering
  \caption{Searches in databases.}
  \label{tab:dbsearches}
  \begin{tabular}{@{}|p{0.32\linewidth}| p{0.60\linewidth}@{}|}
    \hline
    \textbf{Database}&\textbf{Search}\\
    \hline%
    Scopus & TITLE-ABS-KEY ((``software ecosystem*'' OR ``SECO'') AND (``developer experience'' OR ``developer motivation*'' OR ``developer relation*'' OR ``developer satisfaction'' OR ``programmer experience'')) \\ \hline
    Science Direct & Title, abstract, keywords: (``software ecosystem'' OR ``SECO'') AND (``developer experience'' OR ``developer motivation'' OR ``developer relation'' OR ``developer satisfaction'' OR ``programmer experience'')\\ \hline
    IEEE Xplore & (``software ecosystem*'' OR ``SECO'') AND (``developer experience'' OR ``developer motivation*'' OR ``developer relation*'' OR ``developer satisfaction'' OR ``programmer experience'')\\ \hline
    Web of Science & (``software ecosystem*'' OR ``SECO'') AND (``developer experience'' OR ``developer motivation*'' OR ``developer relation*'' OR ``developer satisfaction'' OR ``programmer experience'') (All Fields)\\ \hline
    ACM Digital Library & [[All: ``software ecosystem*''] OR [All: ``seco'']] AND [[All: ``developer experience''] OR [All: ``developer motivation*''] OR [All: ``developer relation*''] OR [All: ``developer satisfaction''] OR [All: ``programmer experience'']]\\ \hline
    Springer Link & (``software ecosystem*'' OR ``SECO'') AND (``developer experience'' OR ``developer motivation*'' OR ``developer relation*'' OR ``developer satisfaction'' OR ``programmer experience'')  \\ \hline
\end{tabular}
\end{table}

\begin{table}
  \scriptsize
  \centering
  \caption{Number of studies per database.}
  \label{tab:dbnumstudies}
  \begin{tabular}{|p{0.65\linewidth} |c|}
     \hline
    \textbf{Database}&\textbf{Search results}\\
    \hline%
    Scopus & 135\\ \hline
    Science Direct & 26\\ \hline
    IEEE Xplore & 5\\ \hline
    Web of Science & 7\\ \hline
    ACM Digital Library & 43\\ \hline
    Springer Link & 114\\ \hline
\end{tabular}
\end{table}

\subsubsection{Step 1.3: Selecting studies}\label{sec:selection}
Below, we list the inclusion (IC) and exclusion (EC) criteria for the studies retrieved by the search string. During the filtering stages, we sought to identify studies that approached DX in SECO.

\begin{itemize}
\item {\textbf{IC1}}: The study approaches DX in SECO.
\item {\textbf{IC2}}: The study presents at least one factor influencing DX in SECO.
\item {\textbf{EC1}}: The study considers DX as the time of experience or seniority.
\item {\textbf{EC2}}: The study is not related to software development in SECO.
\item {\textbf{EC3}}: The study is not primary.
\item {\textbf{EC4}}: The study is not accessible.
\item {\textbf{EC5}}: The study is not a research article or a conference paper.
\item {\textbf{EC6}}: The study is duplicated.
\end{itemize}

Considering these criteria, the study selection has been conducted in three phases:

\begin{itemize}
\item {\textbf{Phase 1 (P1)}}: Read title and abstract.
\item {\textbf{Phase 2 (P2)}}: Read introduction and conclusion.
\item {\textbf{Phase 3 (P3)}}: Read full-text.
\end{itemize}

\begin{figure}[!ht]
  \centering
  \includegraphics[scale=0.7]{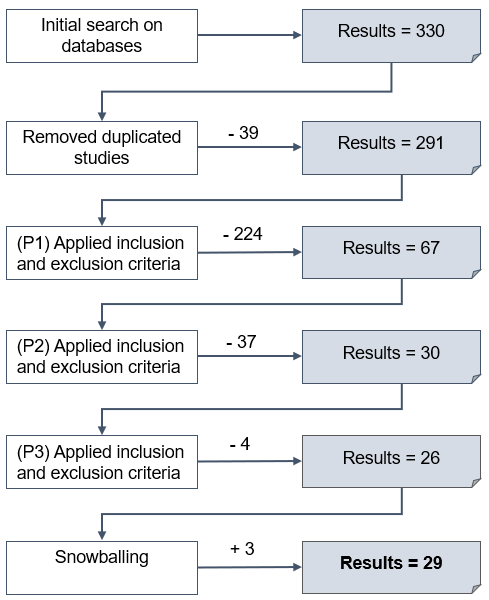}
  \caption{Number of remaining studies in each phase.}
  \label{fig:process}
\end{figure}

This study was conducted in February 2024 with the support of Parsifal\footnote{https://parsif.al/about}, an online tool designed to help researchers perform literature reviews in Software Engineering (SE). We used this tool to remove duplicated studies in an automated way. In each phase, all the criteria were evaluated. To ensure the reliability of the results, two researchers analyzed each study in phases P1 to P3, and they discussed the differences with a third researcher until a consensus was reached. After performing the filtering phases, we applied backward snowballing to verify the selected studies' references and identify more studies to include in this research. At the end of the process, we selected 29 studies for data extraction. Figure~\ref{fig:process} shows the remaining studies in each phase. The snapshots of each phase of our SMS, are available in the supplementary material at~\url{https://doi.org/10.5281/zenodo.13989202}.

\subsubsection{Step 1.4: Extracting data}\label{sec:dataExtraction}
After the snowballing process, the data was extracted and organized in Microsoft Excel\footnote{https://www.microsoft.com/en-us/microsoft-365/excel} spreadsheets. We used the data to present an overview of the primary studies to answer the research questions. The form was structured in the following fields: (i) \textbf{Study ID}: identifier; (ii) \textbf{Study title}: name of the study; (iii) \textbf{Author(s)}: set of names of the authors; (iv) \textbf{Year}: year of publication; (v) \textbf{Country}: country of authors; (vi) \textbf{Venue}: name of publication venue; (vii) \textbf{Factors}: list of factors that influence DX - SQ1; (viii) \textbf{Method(s)}: research method(s) used to find the factors - SQ2; (ix) \textbf{Type(s) of SECO}: characteristics of SECO, such as type of SECO, name, technology etc. - SQ3; and (x) \textbf{Research concern(s)}: issues and challenges frequently highlighted in the literature regarding DX in SECO - SQ4.

\subsubsection{Step 1.5: Analyzing results}\label{sec:analysisResults}
We utilized open and axial coding methods for qualitative analysis to identify and classify DX factors in the literature, as presented by Corbin and Strauss~\cite{corbin2014basics}. During open coding, we carefully read all selected studies, starting from the abstract, introduction, and conclusion sections. The factors were usually clearly stated in those sections; otherwise, we analyzed the remaining sections. To facilitate subsequent reviews and classifications, we transcribed the text excerpts stating the factors, defined their codes, took note of the page, and then classified them into their DX dimension. Table~\ref{tab:coding} presents an example of the form used by the researchers in the coding process.

\begin{table}[!htpb]
\scriptsize
\centering
\caption{Coding form used in identifying DX factors.}
\label{tab:coding}
\begin{tabular}{|p{0.27\linewidth}|p{0.65\linewidth}|}
\hline

\textbf{Study} & S1  \\ 
\hline

\textbf{Factor (code)} & To contribute with new projects  \\ 
\hline

\textbf{Transcribed text} & ``\textit{Several developers associate this factor with innovation, i.e., the creation of innovative applications... This brings new horizons, making some new market opportunities emerge, and a developer can stand out... Contribution with something new helps to get a developer out of the comfort zone so that he/she does not calm down}'' \\ 
\hline

\textbf{Page} & 9  \\ 
\hline

\textbf{DX dimension} & Conative  \\ 
\hline

\end{tabular}
\end{table}

After checking the set of 29 selected studies, we identified 224 candidate factors. We noticed that many of them had similar characteristics related to semantics and context. Therefore, we decided to organize and group them into categories and then merge those that had the same meaning in the context of SECO, following the axial coding method. We intended to make a concise list of DX factors to facilitate their use by researchers and practitioners in SECO scenario analysis.

\begin{figure}[!ht]
    \centering
    \includegraphics[scale=0.9]{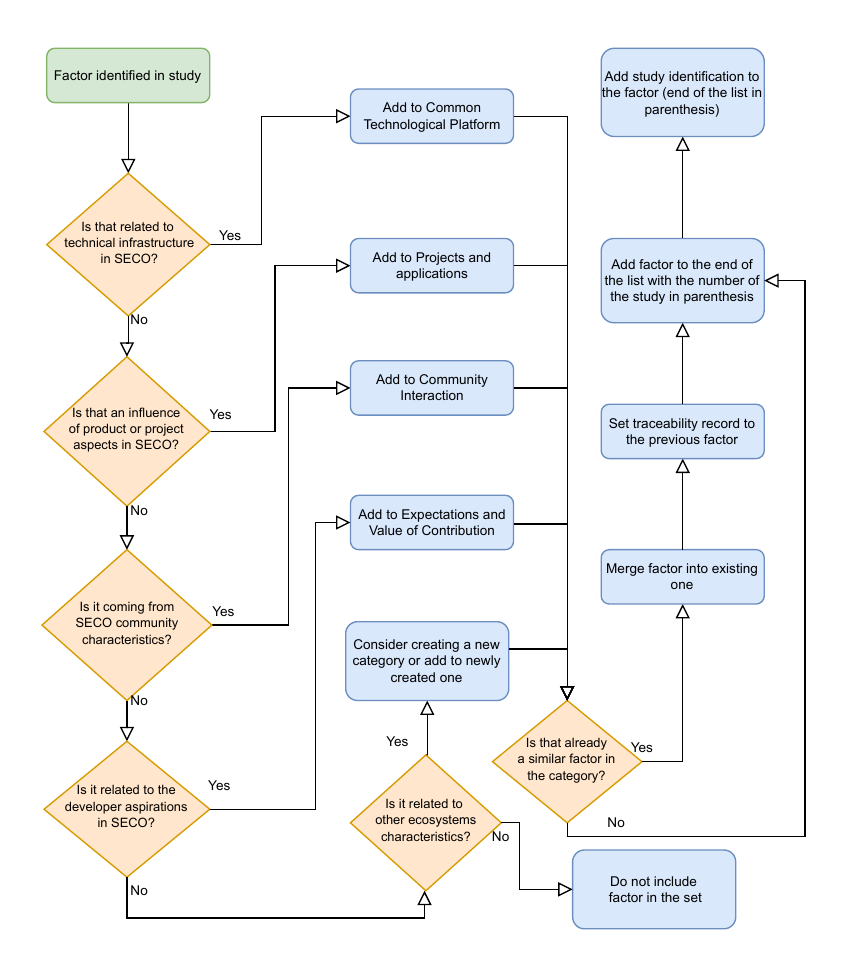}
    \caption{Workflow for identifying and categorizing DX factors.}
    \label{fig:DxProcess}
\end{figure}

Two researchers analyzed each study separately and then came together to compare the texts extracted and codes identified for the categories. If these codes could not help answer the research questions, the researchers discarded them. When there were divergences, they discussed them with a third researcher (expert). We provided a workflow to demonstrate how we made the decisions throughout the analysis in Figure~\ref{fig:DxProcess}. We also make the complete list of candidates, the snapshots of each phase of the SMS, and the final DX factors available in the supplementary material at~\url{https://doi.org/10.5281/zenodo.13989202}.

Given that there is still no consensus on the definition of DX in the context of SE and SECO, we chose to ground it on the definition by Fagerholm and Münch~\cite{Fagerholm2012}, who define it as a broader concept and consider all the experiences that a developer can go through during the software development process. The definitions of the sub-factors in each dimension of this framework inspired us to identify and define the names of the factors presented in this research. Before creating this workflow, we defined four categories inspired by the DX actionable framework of Greiler et al.~\cite{Greiler2022}, which is also based on Fagerholm and Münch~\cite{Fagerholm2012}, but we considered the characteristics of SECO: (i) \textbf{Common Technological Platform}: represents factors related to the technical infrastructure for development provided by a common technological platform of a SECO; (ii) \textbf{Projects and Applications}: represents factors related to the process of developing and distributing applications on a common technological platform of a SECO; (iii) \textbf{Community Interaction}: represents factors related to the interaction between a developer and other developers who integrate a SECO community; and (iv) \textbf{Expectations and Value of Contribution}: represents factors related to expectations and benefits obtained by a developer's contribution in interacting with a common technological platform of a SECO. 

To explain how our decision workflow works, consider the candidate factor ``To contribute with new projects'' (see Table~\ref{tab:coding}). First, we asked the question about the categories. In this case, this factor was related to developers' aspirations in SECO. So, we categorized it into \textbf{Expectations and Value of Contribution}. Then, we analyzed if this factor was similar to another in that category. As it is the first factor, there were no other similarities. Next, we added it to the end of the list in the category and added its study identification in the parenthesis. We replicated this process to all candidate factors and detailed it in the supplementary material.
In the end, we elaborated a list with 27 DX factors.

\subsection{Step 2: DX Factors Evaluation}\label{sec:delphi}
In this step, based on a questionnaire in a Delphi study~\cite{DALKEY1969408}, we evaluated to which extent the list of 27 DX factors, identified in our SMS, influenced third-party software developers to adopt and keep contributing to a SECO. The Delphi study uses an iterative approach to achieve consensus among experts for decision-making, evaluation, or predictive research. It involves collecting and analyzing data anonymously, along with consultations from multiple experts. Their responses are reviewed over a number of rounds until a consensus is reached. Each iteration is similar to a feedback process, which allows and encourages the chosen participants to reassess their judgments provided in previous iterations~\cite{hsu2007delphi,WANG2022100463}. We chose this method because it allows the participants to answer the questionnaire asynchronously and have enough time to reflect on their answers in each round. In our work, we conducted a Delphi study considering the steps presented by Olivero et al.~\cite{OLIVERO2022106874}, as described next.

\subsubsection{Step 2.1: Planning evaluation}\label{sec:planningDelphi}
This step can be divided into the following activities: questionnaire designing, selecting participants, and statistical processing~\cite{OLIVERO2022106874}.

\paragraph{\textbf{Questionnaire designing}}\label{sec:delphiSubject}
Questionnaire designing directly influences the quality of the results and the criteria for choosing the participants~\cite{OLIVERO2022106874}. This activity leads to the identification of research questions that outline the purpose of the questionnaire, along with the creation of the questionnaire itself. Hence, we defined the following SQ5: \textit{``How do third-party developers evaluate the influence of DX factors to adopt and keep contributing to a SECO?''}

Based on the SQ5, we defined the following purpose of the questionnaire: assess experts’ agreement regarding the influence of 27 DX factors identified in the literature on third-party developers adopting and contributing to a SECO. Regarding the questionnaire development, we divided it into three parts. Part I presented the objective of our study and the Informed Consent Form. Part II  had five demographic questions to outline the participants’ profiles. Part III had two mandatory closed questions and an optional open field for comments to evaluate each one of the 27 DX factors. We organize them into four sections according to their respective category.

\begin{figure}[!ht]
    \centering
    \includegraphics[scale=0.7]{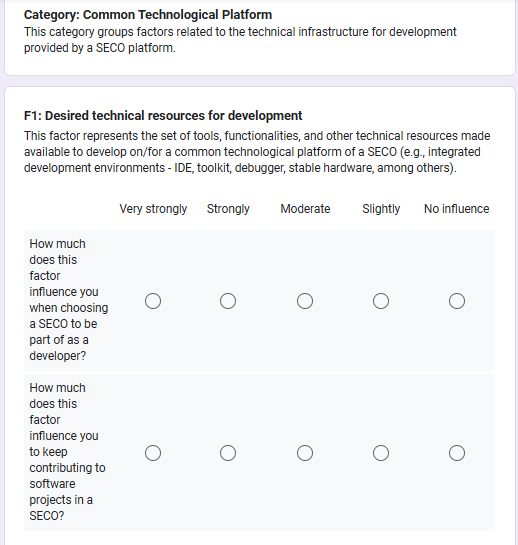}
    \caption{Structure of the questionnaire for evaluating the DX factors in the Delphi study.}
    \label{fig:questionnaire}
\end{figure}

These were the two closed questions in the questionnaire to evaluate each factor: (i) \textit{``How much does this factor influence you when choosing a SECO to be part of as a developer?''} and (ii) \textit{``How much does this factor influence you to keep contributing to software projects in a SECO?''}. To answer both of them, participants could choose between five options of answers inspired by the Likert Scale: very strongly, strongly, moderate, slightly, and no influence. Participants could optionally comment on their choices in the open field. Figure~\ref{fig:questionnaire} illustrates this structure of the questionnaire for evaluating the DX factors in the Delphi study.

We conducted a pilot to evaluate the questionnaire with two researchers with experience in SE and SECO to verify the questionnaire’s clarity and effectiveness. The main suggestions obtained from carrying out the pilot were: (i) provide a specific glossary of SECO terms; (ii) remove the question about age; and (iii) add more examples in the description of factors. After carrying out the pilot, we made the necessary adjustments, and the final version of the questionnaire is available in the supplementary material at~\url{https://doi.org/10.5281/zenodo.13989202}.

The questionnaire was provided to the participants via Google Forms\footnote{https://docs.google.com/forms/}. We kept the panel's composition confidential and communicated with each participant individually to maintain the study's anonymity. We ensured the results were anonymized in each round’s report by removing any references to the identities of the participants.

\paragraph{\textbf{Selecting the participants}}\label{sec:delphParticipants}
Selecting the participants is essential to the success of the Delphi study, as it is directly related to the quality of the results generated~\cite{WANG2022100463, Torrecilla-Salinas2019,AlarabiatRamos2019}. However, identifying suitable participants is one of the main challenges in Delphi studies.  The literature lacks clear guidelines for defining criteria to determine the ideal participant profile~\cite{AlarabiatRamos2019}. Factors such as professional background, expertise in the relevant subject, and willingness to participate can be valid criteria for selecting participants~\cite{Torrecilla-Salinas2019}. Additionally, there is no consensus on the optimal number of participants for a Delphi study~\cite{Torrecilla-Salinas2019}. Many studies suggest that panels typically consist of 15 to 20 members~\cite{WANG2022100463, Torrecilla-Salinas2019,AlarabiatRamos2019}. Torrecilla-Salinas et al.~\cite{Torrecilla-Salinas2019} advised that the number of panelists should be large enough to be representative, yet small enough to keep the process manageable.

As we aimed to evaluate how the list of 27 DX factors influences third-party developers to adopt and keep contributing to a SECO, we defined the population profile for this study as software developers who interact with any SECO to develop their applications. We adopted the non-probabilistic convenience sampling to define the sample due to the impossibility of accurately defining the total number of participants eligible for this research, following the guidelines of Kitchenham et al.~\cite{Kitchenham2015}.

Looking for professionals who fit the defined profile, we sent invitations to third-party developers through email lists of academic and industry organizations and Whatsapp\footnote{https://www.whatsapp.com} groups. Furthermore, we used the snowballing sampling technique, in which the first participants nominated other professionals to participate.

Initially, we invited 50 third-party developers to participate in this study, receiving a positive answer from 21 of them, who forwarded us answers during the different rounds of the method. Table~\ref{tab:DelphiDemographic} presents a summary of information characterizing the profile of the participants. We assigned an identifier (ID) to each software developer, following the order of the interviews carried out (D1 to D21), to identify them throughout this article.

\begin{table*}[!ht]
  \centering\tiny
    \caption{Characterization of the third-party developers.}
    \label{tab:DelphiDemographic}
    \begin{tabular}{@{}| p{0.4cm} | p{1.8cm} | p{1.8cm}  | p{2.8cm} | p{1.8cm} | p{1.5cm} @{}|}
        \hline
        \textbf{ID} & \textbf{Professional Career} & 
        \textbf{Sector} &
        \textbf{Academic Qualification} &
        \textbf{Experience as Professional Developer} &
        \textbf{Types of SECO} \\
        \hline%
 \textbf{D1}\ & Academic & Public & High school or technical degree & 3 years & OSSECO and Hybrid \\\hline
       
 \textbf{D2}\ & Academic & Independent developer & Bachelor's degree & 10 years & OSSECO and Hybrid  \\\hline
        
\textbf{D3}\ & Academic and industrial & Public & High school or technical degree & 4 years & OSSECO and Hybrid \\\hline
        
\textbf{D4}\ & Academic and industrial & Private & High school or technical degree & 3 years & OSSECO   \\\hline
       
\textbf{D5}\ & Academic and Industrial & Private and independent developer & Master's degree & 2 years & PSECO  \\\hline
       
\textbf{D6}\ & Industrial & Private & Specialization degree & 9 years & OSSECO  \\\hline
        
\textbf{D7}\ & Industrial & Private & Specialization degree & 8 years & OSSECO  \\\hline
        
\textbf{D8}\ & Academic and industrial & Public & Master's degree & 2 years & PSECO  \\\hline
       
\textbf{D9}\ & Academic & Public & Bachelor's degree & 2 years & OSSECO and PSECO  \\\hline
 
 \textbf{D10}\ & Academic & Public and independent developer & Bachelor's degree & 3 years & OSSECO  \\\hline
 
 \textbf{D11}\ & Academic & Public & Master's degree & 7 years & OSSECO  \\\hline
 
 \textbf{D12}\ & industrial & Private & Specialization degree & 13 years & OSSECO, PSECO and Hybrid  \\\hline
 
 \textbf{D13}\ & Academic & Private & Bachelor's degree  & 3 years & OSSECO   \\\hline
 
\textbf{D14}\ & Industrial & Public & Specialization degree &  8 years & OSSECO   \\\hline
 
\textbf{D15}\ & Industrial & Private & High school or technical degree & 22 years & OSSECO   \\\hline
 
\textbf{D16}\ & Academic & Public & Bachelor's degree  & 5 years & OSSECO  \\\hline

\textbf{D17}\ & Industrial & Private and independent developer & Specialization degree  & 5 years & OSSECO  \\\hline

\textbf{D18}\ & Industrial & Private & Bachelor's degree  & 2 years & OSSECO and Hybrid  \\\hline

\textbf{D19}\ & Academic & Independent developer & High school or technical degree  & 1 years & OSSECO  \\\hline

\textbf{D20}\ & Industrial & Public & Master's degree  & 3 years & OSSECO \\\hline

\textbf{D21}\ & Academic and industrial & Public and private & Bachelor's degree  & 4 years & OSSECO  \\
 \hline

    \end{tabular}
\end{table*}

Regarding professional careers, eight participants work in an academic career, eight in an industrial career, and five in both academic and industrial careers. Concerning academic qualifications, two participants five a high school or technical degree, seven have a bachelor's degree, five have a specialization degree, and four have a master's degree. The average experience as a professional developer among participants is approximately six years. Finally, regarding the type of SECO with which they interact, thirteen answered OSSECO; four participants answered OSSECO and Hybrid; two answered PSECO; one answered OSSECO, PSECO and Hybrid; and one answered OSSECO and PSECO.

\paragraph{\textbf{Statistical processing}}\label{sec:statiscalExperts}
The statistical processing in Delphi studies involves establishing the statistical methods for analyzing experts' responses in each round, as well as determining the stopping conditions that indicate when further rounds are unnecessary. Therefore, defining the statistical process before conducting the Delphi study prevents authors’ bias in determining the results~\cite{OLIVERO2022106874}. 

The analysis of experts' responses in each Delphi round seeks to determine whether consensus has been reached and to evaluate the stability of responses across successive rounds~\cite{OLIVERO2022106874}. Consensus and stability can be measured using various statistical methods in Delphi studies~\cite{Chalmers2019}. In our Delphi study, we applied descriptive analysis~\cite{thompson_descriptive_2009} to evaluate consensus, applied the Shapiro-Wilk test~\cite{KING2019147} to determine whether the series were normally distributed, and used Pearson and Spearman coefficients~\cite{HaukeKossowski2011} to calculate the correlation between the influence indicators of each DX factor.

Descriptive analysis measures data frequency distribution and central tendency [39]. For the sake of simplicity, we define two influence indicators to calculate correlation. The ``Strong Influence Potential'' (SIP) indicator represents the percentile of subjects who answered that a given factor very strongly or strongly affects their decision to adopt or contribute to a given software ecosystem. Conversely, the ``Weak Influence Potential'' (WIP) indicator represents the percentile of subjects who answered that a given factor either slightly influences or does not influence their decision to adopt or contribute to a given software ecosystem.

The Shapiro-Wilk test aims to assess whether a data distribution approximates a normal distribution. It works with the following hypotheses: (i) null hypothesis: the variable of interest comes from a population with a normal distribution; and (ii) alternative hypothesis: the variable of interest does not come from a population with a normal distribution. The Shapiro-Wilk test produces the W statistic, which will have a p-value associated with it. If the p-value is less than our significance level (usually set to 0.05), we reject the null hypothesis. In other words, we are stating that our data deviates significantly from a normal distribution~\cite{KING2019147}.

We calculated the correlation between the SIP and WIP values over the different factors for adoption and continuing contribution to the ecosystem. Correlation reflects both the strength (indicated by the absolute value of the coefficient) and the direction (indicated by the sign of the coefficient) of the relationship between two variables. The direction can be positive (where an increase in x leads to an increase in y) or negative (where an increase in x results in a decrease in y, or vice versa). There are several types of correlation. In our study, we used Pearson and Spearman coefficients~\cite{HaukeKossowski2011}.

Pearson correlation is a measure of the strength of the linear relationship between two such variables. Spearman’s rank correlation coefficient is a nonparametric (distribution-free) rank statistic proposed as a measure of the strength of the association between two variables~\cite{HaukeKossowski2011}. The correlation coefficient \textit{r} ranges from -1 to 1, and its value indicates different scenarios:

\begin{itemize}
\item If \textit{r} is close to 1, then the variables are linearly positively dependent;
\item If \textit{r} is close to 0, then there is no linear relationship between the variables;
\item If \textit{r} is close to -1, then the variables are linearly negatively dependent.
\end{itemize}

We also tested whether different groups of subjects participating in our study had different opinions concerning the importance of the DX factors. We employed chi-squared tests to determine whether there was a significant association between each DX factor for distinct groups of subjects, according to their background and the type of SECO they are experienced with. A chi-squared test assesses whether the observed frequency of responses about the influence of each DX factor changes according to different groups of subjects, allowing us to address whether, for instance, the most relevant factors change depending on the type of SECO that the subjects are interested in. However, there was no significant difference in the evaluations of the influence of DX factors to adopt and keep contributing to a SECO based on the participants' experience with the different types of SECO, which prevented the creation of new analysis clusters. As a result, we decided to present the overall evaluation results of the DX factors for all types of SECO.

\subsubsection{Step 2.2: Conducting and reporting evaluation}\label{sec:conductingDelphi}
This step comprised the following activities: Round 1, generating statistical data, and further rounds~\cite{OLIVERO2022106874}.

\paragraph{\textbf{Round 1}}\label{sec:delphiRound}
The goal is to introduce the questionnaire to the participants. According to Olivero et al.~\cite{OLIVERO2022106874}, there are two approaches for conducting Round 1 in Delphi studies: using a well-structured questionnaire or a preliminary version for a test round. We opted for the first approach, as we previously conducted a pilot with two researchers experienced in SE and SECO, who assisted us in structuring and reviewing the questionnaire.

\paragraph{\textbf{Generating Statistical Data}}\label{sec:delphiStData}
This stage involves analyzing and interpreting participants’ responses using predefined statistical methods. The organizers of the Delphi study created an anonymous summary of the participants' opinions to evaluate whether the study had met its stopping conditions~\cite{OLIVERO2022106874}.

\paragraph{\textbf{Further Rounds}}\label{sec:delphiFuther}
This process continues until the Delphi study reaches its stopping condition. In each new round, the researchers send the experts a summary of the previous round, which includes statistical data and participants' comments on each question~\cite{OLIVERO2022106874}. Using this summary, the participants can review the overall results and modify or reaffirm their previous answers. Typically, two or three rounds are sufficient to reach a consensus among the expert panel~\cite{OLIVERO2022106874, Torrecilla-Salinas2019}.

\subsubsection{Step 2.3: Determining results and conclusions}\label{sec:delphiConclusions}
Once the stopping condition is met, a report is generated with an analysis of the collected information, and the Delphi study is finished. After completing the evaluation, we generate a ranking of the influence (from most influential to least influential) of DX factors by category, considering the indicator of agreement with the degree of strong influence ($SIP$).

\section{Results}\label{sec:results}
In this section, we present the results from the SMS and the Delphi study. We answer each of the SQ defined in Section~\ref{sec:method}, which together answer our RQ:  \textit{``How do DX factors influence third-party developers to adopt and keep contributing to a SECO?''}

\subsection{Results from the analysis of the state-of-the-art}\label{sec:resultsSMS}

\subsubsection{Demographic Data}\label{demographics}

\begin{figure}[!ht]
\centering
\includegraphics[scale=0.69]{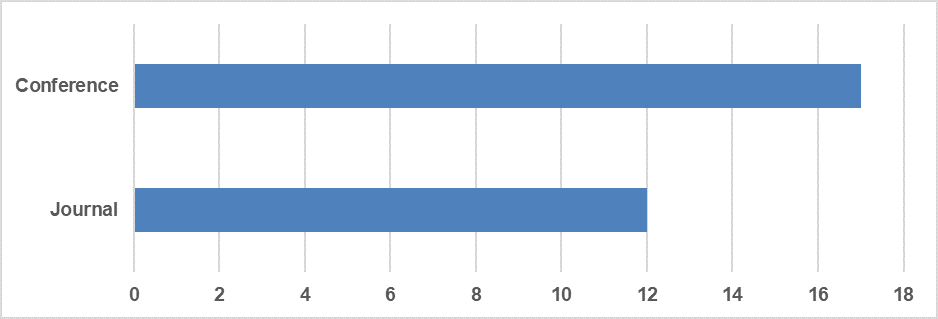}
\caption{Number of studies per venue.}
\label{fig:venue}
\end{figure}

\begin{figure}[!ht]
\centering
\includegraphics[scale=0.69]{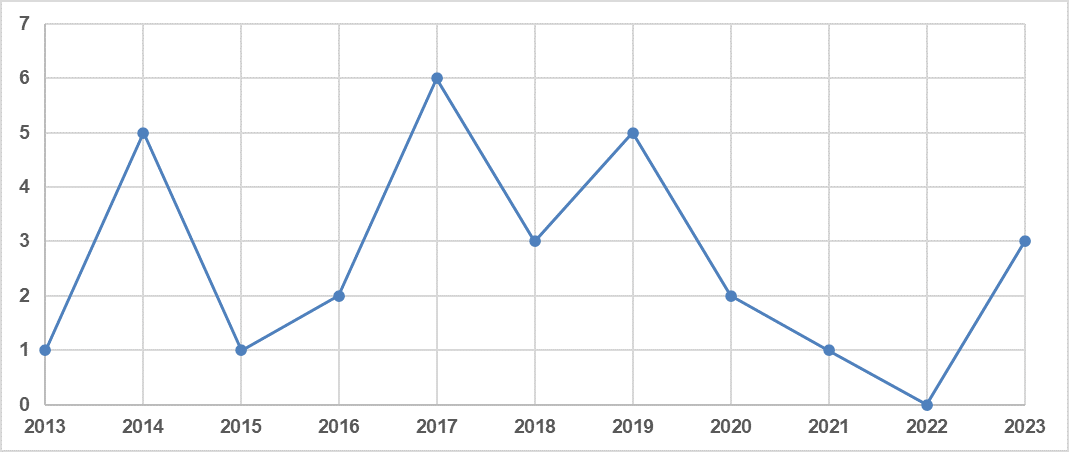}
\caption{Number of studies per year.}
\label{fig:Year}
\end{figure}

The selected studies were published in conferences (17 studies, 59\% of the total), and journals (12 studies, 41\% of the total) as shown in Figure~\ref{fig:venue}. Figure~\ref{fig:Year} shows the distribution of selected studies over the years. We noticed the peak of publications was in 2017 with six studies. Furthermore, it is important to highlight that there were no studies before 2013. This fact can be explained by the publication of the DX conceptual framework in 2012~\cite{Fagerholm2012}, in which the DX concept was clearly defined.

\begin{figure}[!ht]
\centering
\includegraphics[scale=0.8]{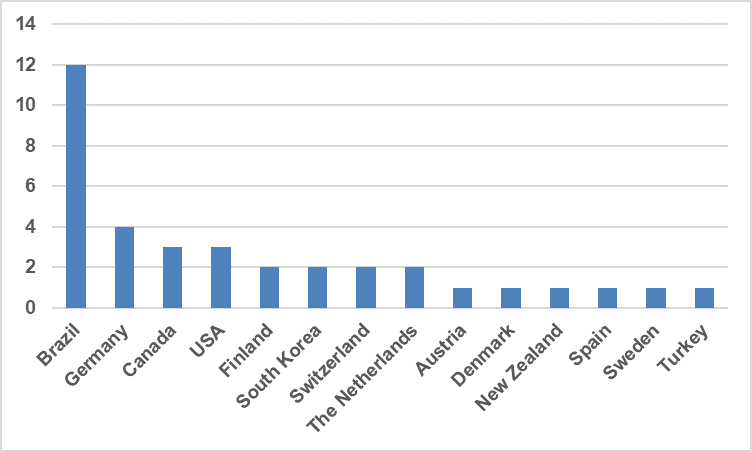}
\caption{Number of studies per country.}
\label{fig:country}
\end{figure}

As noticed in Figure~\ref{fig:country}, relevant primary studies were predominantly conducted in Brazil, with 12 studies, followed by Germany, with 4 studies, and by Canada and the USA, with 3 studies each. Finland, South Korea, Switzerland, and The Netherlands appeared in 2 studies. Finally, Austria, Denmark, New Zealand, Spain, Sweden, and Turkey were present in 1 study each. A study might have researchers from different nationalities, which is why the sum does not correspond to 29 papers. We can highlight that many researchers from Brazil have published works on the subject of DX, mainly involving MSECO.

The complete list of 29 studies can be seen in Table~\ref{tab:studies}, in descending order by year of publication. Identification (ID) codes will be used for referencing studies throughout the next sections. Each study received the identifier (S), followed by a numerical identification (S1-S29). Besides information about the author(s) and year of publication, we added a column to describe which method was used to find the factors and another to describe the type of SECO.

\begin{table*}[!ht]
  \tiny
  \rowcolors{2}{gray!25}{}
  \caption{List of selected studies.}
  \label{tab:studies}
  \begin{tabular}{|c| p{0.44\linewidth} | p{0.11\linewidth}| p{0.15\linewidth}| p{0.1\linewidth} | p{0.03\linewidth}|}
    \hline
   \textbf{ID}&\textbf{Title}&\textbf{Author}&\textbf{Method}&\textbf{Type of SECO}&\textbf{Year}\\
      \hline
    S1 & Factors that affect developers’ decision to participate in a Mobile Software Ecosystem & Steglich et al.~\cite{STEGLICH2023111808} & Literature Review and Interview & Hybrid & 2023\\  \hline
    
    S2 & I didn't find what I wanted - How do Developers Consume Information in Software Ecosystems Portals? & Parracho et al.~\cite{Parracho2023} & Questionnaire & OSSECO, PSECO, and Hybrid & 2023\\  \hline
    
    S3 & Social Networks during Software Ecosystems' Death & Arantes et al.~\cite{Arantes2023} & Data Mining & OSSECO & 2023\\  \hline
    
    S4 & Knowledge Sharing in Digital Platform Ecosystems – A Textual Analysis of SAP’s Developer Community & Kauschinger et al.~\cite{Kauschinger2021} & Case Study and Data Mining & PSECO & 2021\\  \hline
    
    S5 & How do business factors affect developers in mobile software ecosystems? & Steglich et al.~\cite{Steglich2020} & Literature Review and Interview & Hybrid & 2020\\  \hline
    
    S6 & On Value Creation in Developer Relations (DevRel): A Practitioners' Perspective & Fontão et al~\cite{Fontao2020} & Interview & OSSECO and Hybrid & 2020\\  \hline
    
    S7 & Application Developer Engagement In Open Software Platforms: An Empirical Study Of Apple Ios And Google Android Developers & Schaarschmidt et al.~\cite{Schaarschmidt2019} & Questionnaire & Hybrid & 2019\\  \hline
    
    S8 & How do Technical Factors Affect Developers in Mobile Software Ecosystems & Steglich et al.~\cite{Steglich2019} & Interview & Hybrid & 2019\\   \hline
    
    S9 & Perceived input control on online platforms from the application developer perspective: conceptualisation and scale development & Croitor and Belian~\cite{CroitorBenlian2019} & Literature Review, Interview, and Questionnaire & OSSECO, PSECO, and Hybrid & 2019\\   \hline
    
    S10 & Social aspects and how they influence MSECO developers & Steglich et al.~\cite{Steglich2019b} & Literature Review and Interview & Hybrid & 2019\\  \hline
    
    S11 & Why Do People Give Up FLOSSing? A Study of Contributor Disengagement in Open-Source & Miller et al.~\cite{Miller2019} & Questionnaire and Survival Analysis & OSSECO & 2019\\  \hline
    
    S12 & Development as a journey: factors supporting the adoption and use of software frameworks & Myllärniemi et al.~\cite{myllarniemi2018} & Interview and Questionnaire & Hybrid & 2018\\  \hline
    
    S13 & Differential effects of formal and self-control in mobile platform ecosystems: Multi-method findings on third-party developers’ continuance intentions and application quality & Goldbach et al.~\cite{GOLDBACH2018271} & Experiment & Hybrid & 2018\\  \hline
    
    S14 & Mobile Application Development Training in Mobile Software Ecosystem: Investigating the Developer eXperience & Fontão et al.\cite{Fontao2018} & Case Study, Data Mining, and Thematic Analysis & Hybrid & 2018\\  \hline
    
    S15 & A theory of power in emerging software ecosystems formed by small-to-medium enterprises & Valença and Alves~\cite{VALENCA201776} & Case Study, Interview, and Thematic Analysis & PSECO & 2017\\  \hline
    
    S16 & Accommodating openness requirements in software platforms: A goal-oriented approach & Sadi and Yu~\cite{Sadi2017} & Design Case & OSSECO & 2017\\  \hline
    
    S17 & Developer turnover in global, industrial open-source projects: Insights from applying survival analysis & Lin et al.~\cite{Lin2017} & Survival Analysis & OSSECO & 2017\\  \hline
    
    S18 & Entering an ecosystem: The hybrid OSS landscape from a developer perspective & Mäenpää et al.~\cite{Mäenpää2017} & Application of Theory & OSSECO & 2017\\  \hline
    
    S19 & Facing up the primary emotions in Mobile Software Ecosystems from Developer Experience & Fontão et al.~\cite{Fontao2017} & Experiment & Hybrid & 2017\\  \hline
    
    S20 & The Impacts of Mobile Platform Openness on Application Developers' Intention to Continuously Use a Platform: From an Ecosystem Perspective & Choia et al.~\cite{Choia2017} & Questionnaire & Hybrid & 2017\\  \hline
    
    S21 & A power perspective on software ecosystem partnerships & Valença et al.\cite{Valenca2016} & Case Study and Interview & PSECO & 2016\\  \hline
    
    S22 & The social side of software platform ecosystems & De Souza et al.~\cite{deSouza2016} & Interview and Questionnaire & PSECO & 2016\\  \hline
    
    S23 & Designing software ecosystems: How to develop sustainable collaborations? & Sadi et al.\cite{Sadi2015} & Case Study & Hybrid & 2015\\  \hline 
    
    S24 & Acceptance of monetary rewards in open-source software development & Krishnamurthy et al.\cite{KRISHNAMURTHY2014632} & Questionnaire & OSSECO & 2014\\  \hline
    
    S25 & Bridges and barriers to hardware-dependent software ecosystem participation – A case study & Wnuk et al.\cite{WNUK20141493} & Case Study and Interview & PSECO & 2014\\  \hline
    
    S26 & Factors affecting application developers’ loyalty to mobile platforms & Ryu et al.~\cite{RYU201478} & Questionnaire & Hybrid & 2014\\  \hline
    
    S27 & Joining a smartphone ecosystem: Application developers’ motivations and decision criteria & Koch and Kerschbaum~\cite{KOCH20141423} & Literature review, Interview, and Questionnaire & Hybrid & 2014\\  \hline
    
    S28 & Software engineering beyond the project – Sustaining software ecosystems & Dittrich~\cite{DITTRICH20141436} & Interview and Ethnographical Research & PSECO & 2014\\  \hline
    
    S29 & Categorizing developer information needs in software ecosystems & Haenni et al.\cite{Haenni2013} & Interview & OSSECO, PSECO, and Hybrid & 2013\\
    \hline
\end{tabular}
\end{table*}

After screening selected primary studies in this SMS, collected data were analyzed to answer each sub-question (SQ). The raw data and all the steps necessary to reproduce the results are detailed in the supplementary material.

\subsubsection{SQ1: Which factors affect DX in SECO?}\label{sec:factors}

\begin{figure}[!ht]
    \centering
    \includegraphics[scale=0.43]{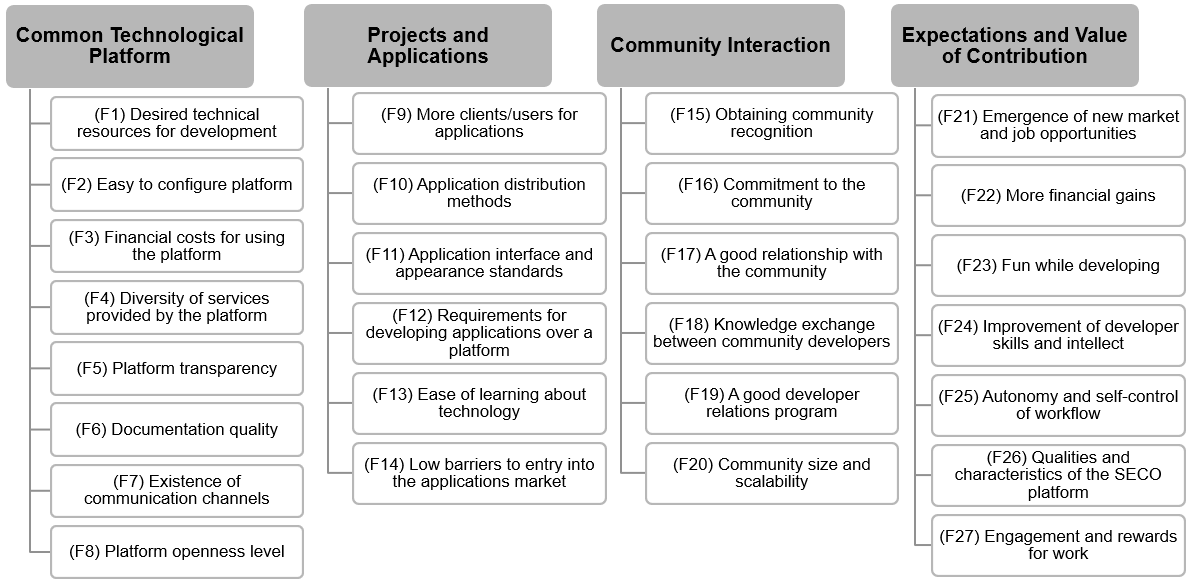}
    \caption{List of DX factors identified in our SMS.}
    \label{fig:ListDxFactors}
\end{figure}

To answer this SQ, we aimed to gather the most relevant factors which influence DX in SECO. Next, we grouped them into four categories based on the framework proposed by Greiler et al. \cite{Greiler2022} and elaborated the schema seen in Figure~\ref{fig:ListDxFactors}. It lists the DX factors in four categories in order: \textbf{Common Technological Platform}, \textbf{Projects and Applications}, \textbf{Community Interaction}, and \textbf{Expectations and Value of Contribution}.

In Common Technological Platform, we can find factors related to the technical infrastructure for development provided by a common technological platform of a SECO. The factor~\textbf{(F1) Desired technical resources for development} represents the set of tools, functionalities, and other technical resources made available to develop on/for a common technological platform of a SECO (e.g., integrated development environments - IDE, toolkit, debugger, stable hardware, among others) (S1, S4, S6, S8, S23, S25, S26, and S27). The factor \textbf{(F2) Easy to configure platform} represents the ease of configuring the technical resources made available and preparing the development environment to contribute to a common technological platform of a SECO (S1, S5, S8, S12, S14, S19, and S25). The factor\textbf{ (F3) Financial costs for using the platform} represents the financial costs (fees, subscriptions, and/or licenses) involved in using a common technological platform of a SECO (S1 and S5). The factor \textbf{(F4) Diversity of services provided by the platform} represents the set of services and capabilities provided by a common technological platform to support software development (e.g., technical support, customization, good hardware performance, integrations, maintenance, application programming interfaces - API, among others) (S1, S5, S6, S8, S12, S25, S27, and S29).

In the same category, we also have the factor \textbf{(F5) Platform transparency}, which represents the set of characteristics that allow the developer access, ease of use, quality of content, understanding, and auditing of information and processes related to a common technological platform (e.g., documentation, communication channels, organization's mission, quality criteria, among others) (S2, S6, S12, S19, S25, and S29). The factor \textbf{(F6) Documentation quality} represents the set of characteristics that promote the quality of the technical documentation made available by a common technological platform so that developers can know and learn how to contribute to a given SECO (e.g., correctness, clarity, integrity, timeliness, among others) (S2, S6, S12, S22, S25, S28, and S29). The factor \textbf{(F7) Existence of communication channels} represents the communication channels that allow the software developer to interact with other developers and/or with employees of the organization that manages the SECO common technological platform (e.g., forums, issues, emails, communities, among others) (S2, S3, S7, S15, S25, and S28). Finally, the factor \textbf{(F8) Platform openness level} represents the level of openness of a common technological platform for access, modification and distribution of its technical resources freely by third-party developers (e.g., source code, tools, among others) (S7, S20, S23, S25, and S27).

In Projects and Applications, we are focusing on factors related to the process of developing and distributing applications on a common technological platform of a SECO. The factor \textbf{(F9) More clients/users for applications} represents how much developing with technology from a given SECO can increase the number of customers/users for a developer's applications (e.g., user demand, advertising, compatibility with different devices, among others) (S1, S5, S20, S22, S25, and S26). The factor \textbf{(F10) Application distribution methods} represents the diversity of ways of distributing applications developed by external developers, considering the number of distribution channels, SECO' positioning in the market, market size, user feedback, among others (S1, S5, S22, and S27). The factor \textbf{(F11) Application interface and appearance standards} represents the existence of interface and appearance standards that must be adopted in the development of applications on the common technological platform, aiming at creating an identification mark and maintaining the internal and external quality of the applications (S8, S25, and S28).

In the same category, we also have the factor \textbf{(F12) Requirements for developing applications over a platform}, which represents the requirements imposed by a keystone so that third-party developers can develop on a common technological platform (e.g., regulatory requirements or policies, minimum technical requirements, implementation quality, architectural and/or interface standards for applications, among others) (S9, S25, S28, and S29). The factor \textbf{(F13) Ease of learning about technology} represents the degree of ease in learning how to use SECO platform technology to develop applications, from the third-party developer's point of view (S2, S7, S14, S17, S22, S23, S26, and S27). Finally, the factor \textbf{(F14) Low barriers to entry into the applications market} considers common SECO technological platforms that present low barriers to entry for new developers in the application development and trading market (e.g., technical compatibility, fee collection, among others) (S1, S6, S9, S16, S23, and S27).

In Community Interaction, we group some aspects related to the interaction between a developer and other developers who integrate a SECO community. The factor \textbf{(F15) Obtaining community recognition} considers the developer's expectation that their work and/or contributions will be recognized by the SECO community (e.g., gaining reputation, promoting their contributions, recognizing expertise, among others) (S1, S6, S10, S14, S21, and S23). The factor \textbf{(F16) Commitment to the community} considers the motivations, ideologies, and identification of the developer with other developers and with the actions of a SECO community (e.g., contributing to open-source projects, as they believe in this software development philosophy) (S1, S10, S24, and S28). The factor \textbf{(F17) A good relationship with the community} considers the construction of good relationships between the developer and other members of a SECO community (e.g., sense of community, solid relationships, robust network, good communication, broad and fair community, among others) (S1, S2, S3, S14, S18, and S28).

In the same category, we also have the factor \textbf{(F18) Knowledge exchange between community developers}, which considers the exchange of knowledge and expertise between developers in a SECO community to help each other (e.g., recommendations, interaction between peers, solving problems or doubts, learning from more experienced developers, among others) (S1, S10, S12, S18, S22, and S28). The factor \textbf{(F19) A good developer relations program} considers the developer relations program promoted by the keystone of a SECO, where developers of a common technological platform (DevRel) offer support, training, and feedback, bringing the developer community closer to the organization and the product/service (S6, S14, S22, and S28). Finally, the factor \textbf{(F20) Community size and scalability} considers the activity level, size, and growth prospects of a SECO community over time (e.g., the developer opts for communities that have a growth trend in the short or medium term) (S6, S10, S12, S16, and S27).

In Expectations and Value of Contribution, we put together some factors related to expectations and benefits obtained by a developer's contribution in interacting with a common technological platform of a SECO. The factor \textbf{(F21) Emergence of new market and job opportunities} considers the activity level, size, and growth prospects of a SECO community over time (e.g., a developer opts for communities that have a growth trend in the short or medium term) (S1, S5, S10, S11, S22, and S25). The factor \textbf{(F22) More financial gains} considers that the developer expects to obtain greater financial gains by interacting with a SECO platform through application sales, contribution rewards, and market demand (S1, S5, S6, and S23). The factor \textbf{(F23) Fun while developing} considers that the developer expects to have fun while developing on/for a SECO platform (S1, S10, S17, S23, S24, and S27).

In the same category, we also have the factor \textbf{(F24) Improvement of developer skills and intellect}, which considers that the developer expects to improve his skills and intellect when interacting with a SECO platform (e.g., improve programming skills, creativity, innovation and/or learn new skills) (S1, S6, S10, S17, S23, and S27). The factor \textbf{(F25) Autonomy and self-control of workflow} considers that a developer expects a keystone to promote conditions for autonomy and self-control of the workflow when integrating with a SECO platform (S1, S10, S11, S13, and S17). The factor \textbf{(F26) Qualities and characteristics of the SECO platform} considers that a developer expects to identify qualities and characteristics of the SECO platform that motivate him to contribute (e.g., credibility, reputation of keystone, scalability, recognition, and appreciation of developers' contributions, among others) (S5, S6, S10, S15, S18, S21, and S25). Finally, the factor \textbf{(F27) Engagement and rewards for work} considers that the developer expects to find motivations to engage and obtain rewards when interacting with a SECO platform (e.g., sense of accomplishment, feeling of involvement, working on popular projects, obtaining satisfactory returns, ideological or altruistic motivations, meeting expectations, among others) (S3, S6, S11, S14, S17, S18, S21, S24, S25, and S26).

\subsubsection{SQ2: Which research methods are used to obtain these factors?}\label{sec:researchMethods}

\begin{figure}[!ht]
\centering
\includegraphics[scale=0.69]{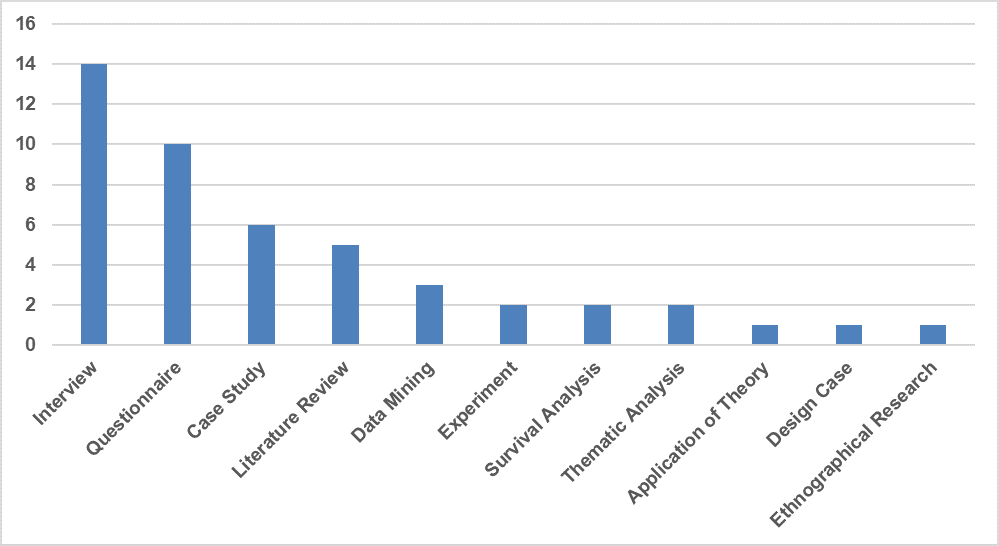}
\caption{Research methods used to obtain DX factors in the studies.}
\label{fig:methodsSMS}
\end{figure}

The research methods applied for finding the factors are presented in Figure~\ref{fig:methodsSMS}. The most used method is Interview (S1, S5, S6, S8, S9, S10, S12, S15, S21, S22, S25, S27, S28, and S29), followed by Questionnaire (S2, S7, S9, S11, S12, S20, S22, S24, S26, and S27). Since this topic is more related to qualitative analysis, this was an expected result. We believe that this happens because the evaluation of the experience is something subjective and influenced by several variables and each subject can have his/her perceptions. Thus, these two methods help researchers to better understand these nuances.

Moreover, we can highlight that interviews allow understanding each factor deeper, going into the details of each aspect of it. Although it is more difficult to go over these details for surveys with questionnaires, it can reach more developers also due to the characteristics of geolocation. The other methods appeared more isolated in some studies but also brought valid and relevant results, most of them also related to qualitative analysis.

\subsubsection{SQ3: Which types of ecosystems have been considered?}\label{sec:typesSECO}

Table~\ref{tab:typesSECO} shows the types of ecosystems approached by the selected studies. Hybrid SECO is the type of ecosystem most considered by the studies (S1, S2, S5, S6, S7, S8, S9, S10, S12, S13, S14, S19, S20, S23, S26, S27, and S29), following by OSSECO (S2, S3, S6, S9, S11,S16, S17, S18, S24, and S29), and PSECO (S2, S4, S9, S15, S21, S22, S25, S28, and S29).

\begin{table}[!htpb]
\scriptsize
\centering
\caption{Types of ecosystems considered in the studies.}
\label{tab:typesSECO}
\begin{tabular}{|p{0.4\linewidth}|c|}
\hline

\textbf{Type of SECO} & \textbf{Percentage}  \\ 
\hline

Hybrid & 47\% \\ 
\hline

OSSECO & 28\%  \\ 
\hline

PSECO & 25\% \\ 
\hline

\end{tabular}
\end{table}

Concerning the names of the ecosystem technologies considered in the studies, we have: (i) Hybrid: Android, iOS, and Windows Phone; (ii) OSSECO: Linux, Unity, AngularJS, PhantomJS, Moment.js, Twitter, Chrome, and Firefox; and (iii) PSECO: Axis, Microsoft, SIM, and SAP.

\subsubsection{SQ4: Which concerns are pointed out for DX in SECO? }\label{sec:concerns}

\vspace{8px}
\noindent
\textbf{Attracting or maintaining developers in SECO over time}: this is one of the main indicators that affect the health and sustainability of SECO. Third-party developers help a keystone to expand its SECO and consolidate itself in the market over time. Thus, it is essential to define processes to attract and engage more actors, mainly developers. In this situation, promoting actions that provide a satisfactory DX, decreasing entry barriers for new developers, and increasing interest and engagement in a common technological platform (S2, S3, S6, S7, S12, S13, S14, S17, S18, S19, S21, S23, and S25).

\vspace{8px}
\noindent
\textbf{Ensuring SECO sustainability}: SECO sustainability has been based on two main pillars: i) adaptation to new technologies, resources, and trends; and ii) attraction and retention of members in the SECO community. As in the previous concern, this ability to attract new developers to an ecosystem is crucial for its sustainability and its survival. Therefore, it is necessary to explore both pillars, studying the social side and DX of software ecosystems, as it can shed light on aspects that can influence their sustainability (S5, S8, S9, S10, S21, S22, and S24).

\vspace{8px}
\noindent
\textbf{Defining the level of SECO openness}: there are large volumes of data transiting through SECO, due to its open characteristic that allows contributions from different actors who can enter and leave the ecosystem at any time. Often, keystones face problems related to access and use of this data by them. Although on the one hand, free access to data can be positive from the perspective of transparency, on the other hand, undefined access to a large volume of data can make it difficult to interact with a common technological platform. Therefore, it is necessary to build visual data tools that add transparency resources and investigate how to moderate the permissions and access levels of different actors to data shared in SECO, contributing to better usability and DX on using the platform (S6, S11, S16, and S20).

\vspace{8px}
\noindent
\textbf{Dealing with power relations within SECO}: as there are different actors, with different interests, interacting with the common technology platform, conflicts of interest may arise that must be managed by the keystone. In some situations, power is exercised strongly and forcefully by a company or group of actors that retains control of valuable assets and, therefore, can impose its needs on players or developers, potentially causing an imbalance within the ecosystem. Therefore, to survive and grow in the so-called ``ecosystem war'', it is important that platform providers understand how they can create and maintain a positive relationship between the different actors, especially the developers (S9, S15, S26, and S28).

\subsection{Results of the DX Factors evaluation}\label{sec:resultsDelphi}

The DX factors evaluation occurred through a Delphi study with 21 third-party developers who interact with any SECO to develop their applications. We carried out two rounds to reach a consensus among the participants. The study questions were measured on a 5-point Likert scale to represent the factor influence: very strongly, strongly, moderate, slightly, no influence. In this study, we aimed to answer SQ5: \textit{``How do third-party developers evaluate the influence of DX factors to adopt and keep contributing to a SECO?''}
 
\subsubsection{Round 1}\label{sec:resultsDelphiR1}

Round 1 of the Delphi study was conducted in September 2024. The participants evaluated the 27 DX factors identified in the literature. They answered two questions in the questionnaire to each factor: (i) \textit{``How much does this factor influence you when choosing a SECO to be part of as a developer?''} and (ii) \textit{``How much does this factor influence you keep contributing to software projects in a SECO?''}. Participants could optionally comment on their choices in an open field.

\begin{figure}[!ht]
    \centering
    \includegraphics[width=1.0\linewidth]{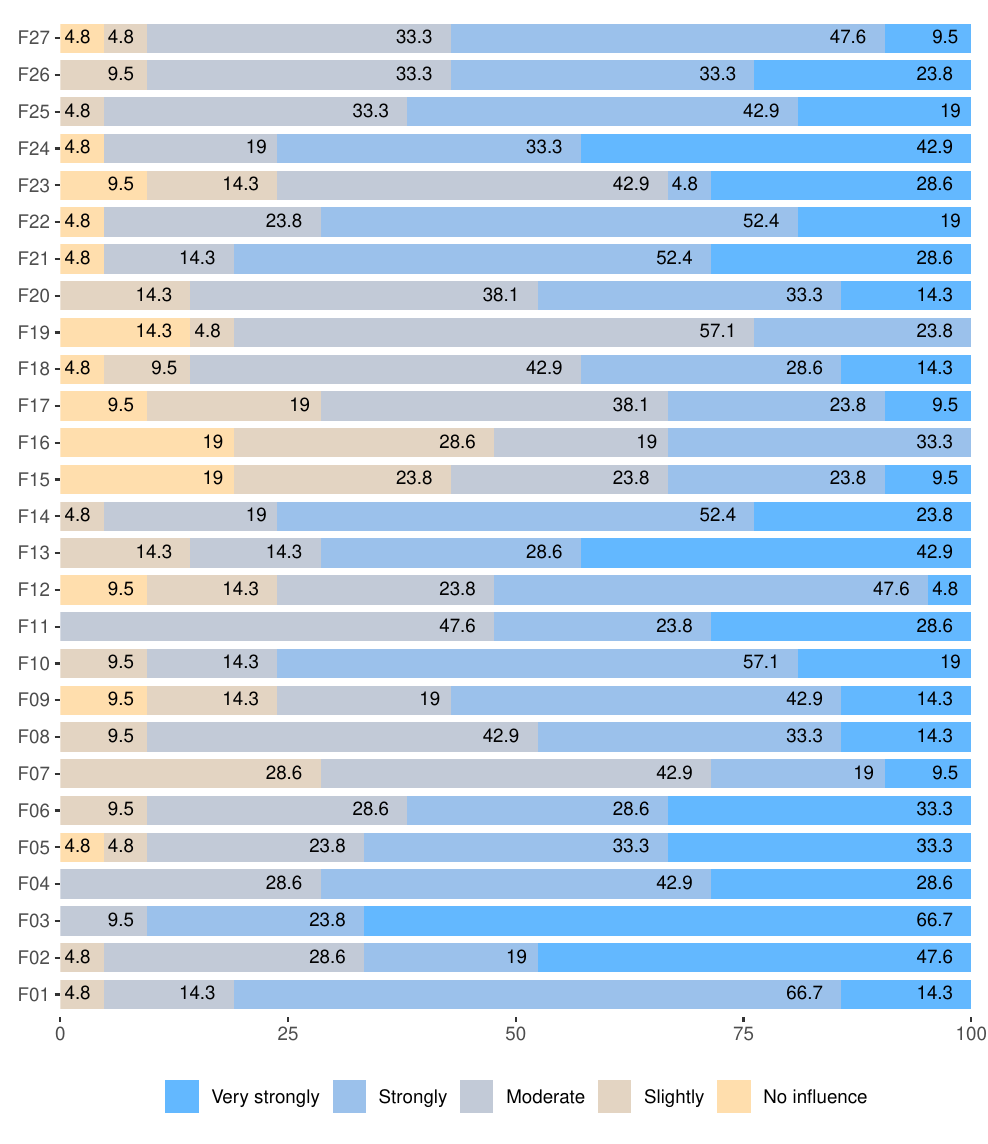}
    \caption{Frequency analysis for the factors under analysis and their influence on selecting a software ecosystem.}
    \label{fig:escolha-fatores}
\end{figure}

We received 21 responses for the proposed survey. Figure \ref{fig:escolha-fatores} presents the distribution of the selected factors' influence on a subject choosing a given ecosystem to start developing software. Each factor is represented in a row. Each row has three to five blocks that must be read from left to right to increase the influence level for the related factor. Numbers in the bars indicate the percentile of respondents that answered the question associated with the specific factor for the related influence level. For instance, 4.8\% of the subjects answered that factor F1 (\textit{Desired technical resources for development}) only slightly influenced their decision in selecting a software ecosystem, while 66.7\% respondents answered that the same factor strongly affected their decision on developing for a given ecosystem.

\begin{figure}[!ht]
    \centering
    \includegraphics[width=1.0\linewidth]{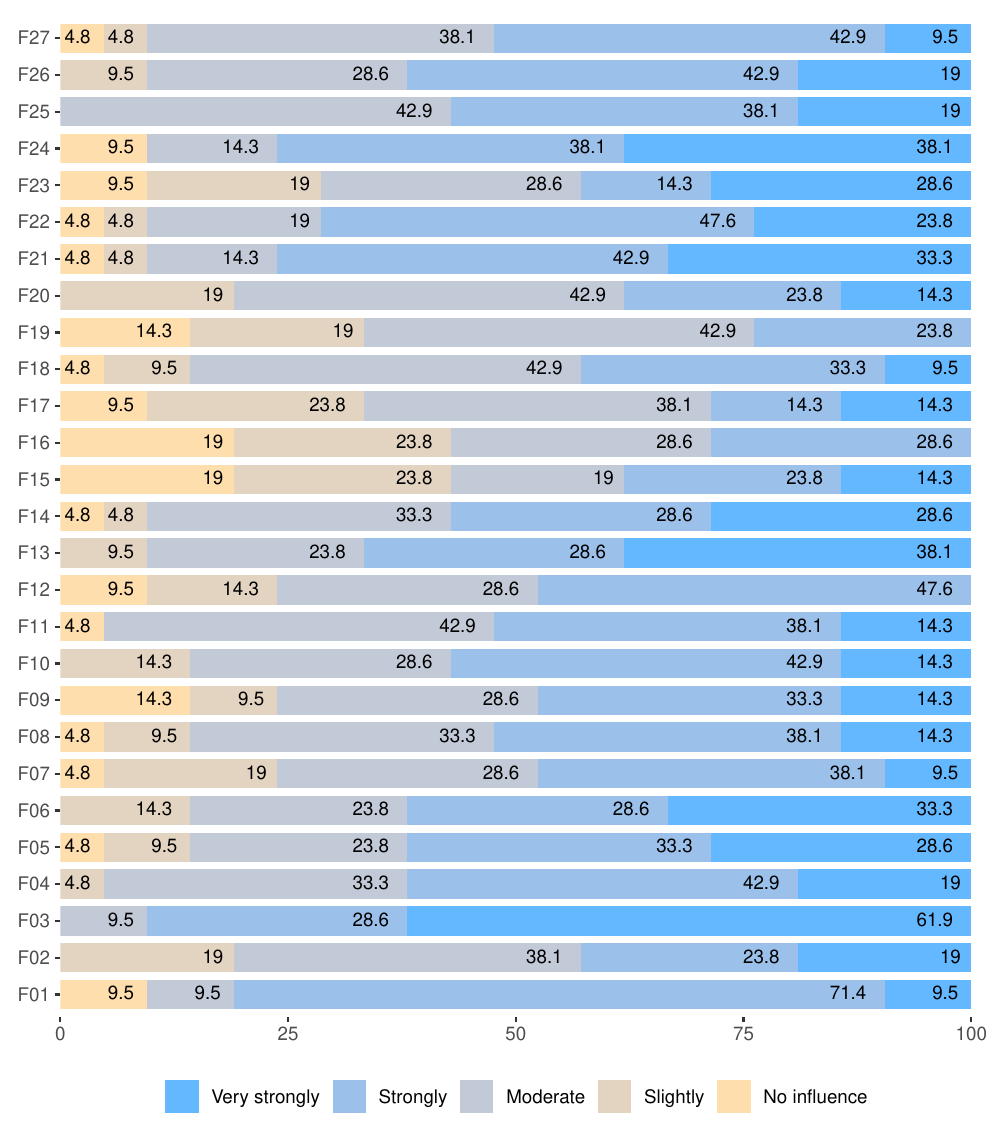}
    \caption{Frequency analysis for the factors under analysis and their influence on continuing to contribute to a software ecosystem.}
    \label{fig:manutencao-fatores}
\end{figure}

Figure \ref{fig:manutencao-fatores} presents the distribution of the selected factors' influence on a subject's intention to keep contributing to a given software ecosystem. As in the former figure, each factor is represented in a row with three to five blocks representing increasing influence levels for the related factor from left to right. Numbers in the bars indicate the percentile of respondents that answered the question associated with the specific factor for the related influence level.

As we mentioned in Section~\ref{sec:planningDelphi}, the SIP indicator represents the percentile of subjects who answered that a given factor strongly or very strongly influences their decision to adopt or contribute to a given software ecosystem. Conversely, the WIP indicator represents the percentile of subjects who answered that a given factor either slightly influences or does not affect their decision to adopt or contribute to a given software ecosystem. Table \ref{tab:indicators} presents the values for these indicators for each factor, separating the adoption of a software ecosystem from the continuing contribution to the ecosystem.

\begin{table}[!ht]
\centering\scriptsize
\caption{Ecosystem adoption and continued contribution influence indicator values for each factor.}
\begin{tabular}{|l|rr|rr|}
\hline
\multirow{2}{*}{\textbf{Factor}} & \multicolumn{2}{|c|}{Adoption} & \multicolumn{2}{|c|}{Keep} \\
\cline{2-5}
& \textbf{SIP} & \textbf{WIP} & \textbf{SIP} & \textbf{WIP} \\
\hline
F1 & 81.0 &  4.8 & 81.0 &  9.5 \\\hline
F2 & 66.7 &  4.8 & 42.9 & 19.0 \\\hline
F3 & 90.5 &    - & 90.5 &    - \\\hline
F4 & 71.4 &    - & 61.9 &  4.8 \\\hline
F5 & 66.7 &  9.5 & 61.9 & 14.3 \\\hline
F6 & 61.9 &  9.5 & 61.9 & 14.3 \\\hline
F7 & 28.6 & 28.6 & 47.6 & 23.8 \\\hline
F8 & 47.6 & 9.52 & 52.4 & 14.3 \\\hline
F9 & 57.1 & 23.8 & 47.6 & 23.8 \\\hline
F10 & 76.2 &  9.5 & 57.1 & 14.3 \\\hline
F11 & 52.4 &    - & 52.4 &  4.8 \\\hline
F12 & 52.4 & 23.8 & 47.6 & 23.8 \\\hline
F13 & 71.4 & 14.3 & 66.7 &  9.5 \\\hline
F14 & 76.2 &  4.8 & 57.1 &  9.5 \\\hline
F15 & 33.3 & 42.9 & 38.1 & 42.9 \\\hline
F16 & 33.3 & 47.6 & 28.6 & 42.9 \\\hline
F17 & 33.3 & 28.6 & 28.6 & 33.3 \\\hline
F18 & 42.9 & 14.3 & 42.9 & 14.3 \\\hline
F19 & 23.8 & 19.0 & 23.8 & 33.3 \\\hline
F20 & 47.6 & 14.3 & 38.1 & 19.0 \\\hline
F21 & 81.0 &  4.8 & 76.2 &  9.5 \\\hline
F22 & 71.4 &  4.8 & 71.4 &  9.5 \\\hline
F23 & 33.3 & 23.8 & 42.9 & 28.6 \\\hline
F24 & 76.2 &  4.8 & 76.2 &  9.5 \\\hline
F25 & 61.9 &  4.8 & 57.1 &    - \\\hline
F26 & 57.1 &  9.5 & 61.9 &  9.5 \\\hline
F27 & 57.1 &  9.5 & 52.4 &  9.5 \\\hline

\end{tabular}
\label{tab:indicators}
\end{table}

We notice that most factors have stronger than weaker influence potential, i.e., most subjects believe that the large majority of the proposed factors have a strong influence both on their decision to adopt an ecosystem and to keep developing for it. Interesting exceptions are F15 (\textit{Obtaining community recognition}) and F16 (\textit{Commitment to the community}) which are less important for both decisions, while F17 (\textit{A good relationship with the community}) and F19 (\textit{A good developer relations program}) are less relevant to keep a developer engaged in a given ecosystem. Regarding factors F15 and F16, participant D3 commented that he recognizes the importance of commitment to the community, but that personally, it did not influence him as much: \textit{``I believe that contributing to an OSSECO is important for the software development community, but personally it doesn't influence me that much''}. Concerning F17 and F19, participants did not comment on the reason why they are less relevant to keep a developer engaged in a given ecosystem. D8 highlighted a positive aspect of F19: \textit{``An active DevRel program that provides support, training, and feedback not only facilitates my learning but also creates a bond between the community and the organization. This connection increases my confidence and motivation to contribute to projects, as I feel supported and valued. If this program is weak or non-existent, my willingness to participate and develop on the platform tends to be significantly reduced''}.

Next, we calculated correlations between the $SIP$ and $WIP$ values over the different factors for adoption and continuing contribution to the ecosystem. We applied a Shapiro-Wilk test to determine whether the series were normally distributed and found that both SIP series were close to a bell-shaped distribution while the WIP series had heavy left tails. Therefore, we used the Pearson coefficient to calculate the correlation between the $SIP$ series and the Spearman coefficient for the $WIP$ series. Both coefficients presented a strong correlation (0.88 for $SIP$ and 0.87 for $WIP$). Therefore, we will analyze only the adoption indicators in the further rounds, since the distribution of the continuing development indicators is closely related to the former one.

We studied whether there was agreement or disagreement among groups of developers concerning the influence of the proposed factors. For that purpose, we applied chi-squared tests over different groups of subjects. Starting from their formation, we found that different subject groups agreed on the influence of factors F8 (\textit{Platform openness level}), F9 (\textit{More clients/users for applications}), F12 (\textit{Requirements for developing applications over a platform}), F16 (\textit{Commitment to the community
}), and F20 (\textit{Community size and scalability}) (i.e., there was not a significant difference among the WIP and SIP series for these groups at 95\% certainty).

Next, taking into consideration the kind of ecosystem on which subjects have experience, we found subjects working on different types of ecosystems agreed on the influence of sixteen factors: F1, F2, F6, F9, F10, F12, F13, F14, F15, F18, F20, F21, F22, F23, F24, and F27.
F11 (\textit{Application interface and appearance standards}) is considered more important by open-source developers (SIP = 47.6\%) than by developers of hybrid (9.5\%) or proprietary ecosystems (14.3\%). Regarding this factor, D8 stated: \textit{``This factor is important to me when choosing a SECO. Following standards helps to create a consistent visual identity and ensure the quality of applications. This not only improves the user experience, but also facilitates collaboration with other developers. When a SECO has clear interface guidelines, I am more motivated to contribute, as I know that my work will align with a recognizable and well-structured brand. This brings trust and professionalism to the project.''} 

F16 (\textit{Commitment to the community}) and F17 (\textit{A good relationship with the community}) have little or no influence over developers of open-source or hybrid ecosystems, while these factors are highly regarded by proprietary ecosystem developers. F25 (\textit{Autonomy and self-control of workflow}) and F26 (\textit{Qualities and characteristics of the SECO platform}) have moderate influence over the decision of hybrid and proprietary ecosystem developers but are very important for open-source developers. F4 (\textit{Diversity of services provided by the platform}) and F5 (\textit{Platform transparency}) are also very important for open-source developers. Concerning F5, D13 commented: \textit{``In my opinion, platform transparency is one of the crucial factors for the good development of a project. Since these artifacts store all the formal knowledge (technical and social) of the platform, developers must keep them updated for the inclusion and training of new users and good project management''}. 

F3 (\textit{Financial costs for using the platform}) is important for all developers, but particularly important for open-source developers. About this factor, D18 stated: \textit{``Normally the best ones will be more expensive, but there must be a balance between need and practicality. The use of a more expensive platform is due to the scope of the project and the need for its use later, with larger projects or several smaller projects that benefit from its use''}. F7 (\textit{Existence of communication channels}), F8 (\textit{Platform openness level}), and F19 (\textit{A good developer relations program}) are moderately important for open-source developers, while their importance for other developers is smaller.


\subsubsection{Round 2}\label{sec:resultsDelphiR2}
Round 2 of the Delphi study was conducted in October 2024. Participants again assessed the 27 DX factors identified in the literature, but this time they could view the total score from Round 1 and comments from other participants before answering the questionnaire. As we mentioned in the analysis of Round 1, the coefficients of the two questions showed a strong correlation (0.88 for $SIP$ and 0.87 for $WIP$). Therefore, we chose to ask only one question for each factor in Round 2, because the responses would be representative of both situations described in Round 1 (adopting and continuing to contribute to a SECO). In this round, we asked the following question: \textit{``How much does this factor influence you when choosing a SECO to be part of as a developer?''}. Participants could also optionally comment on their choices in an open field.

\begin{figure}[!ht]
    \centering
    \includegraphics[scale=0.8]{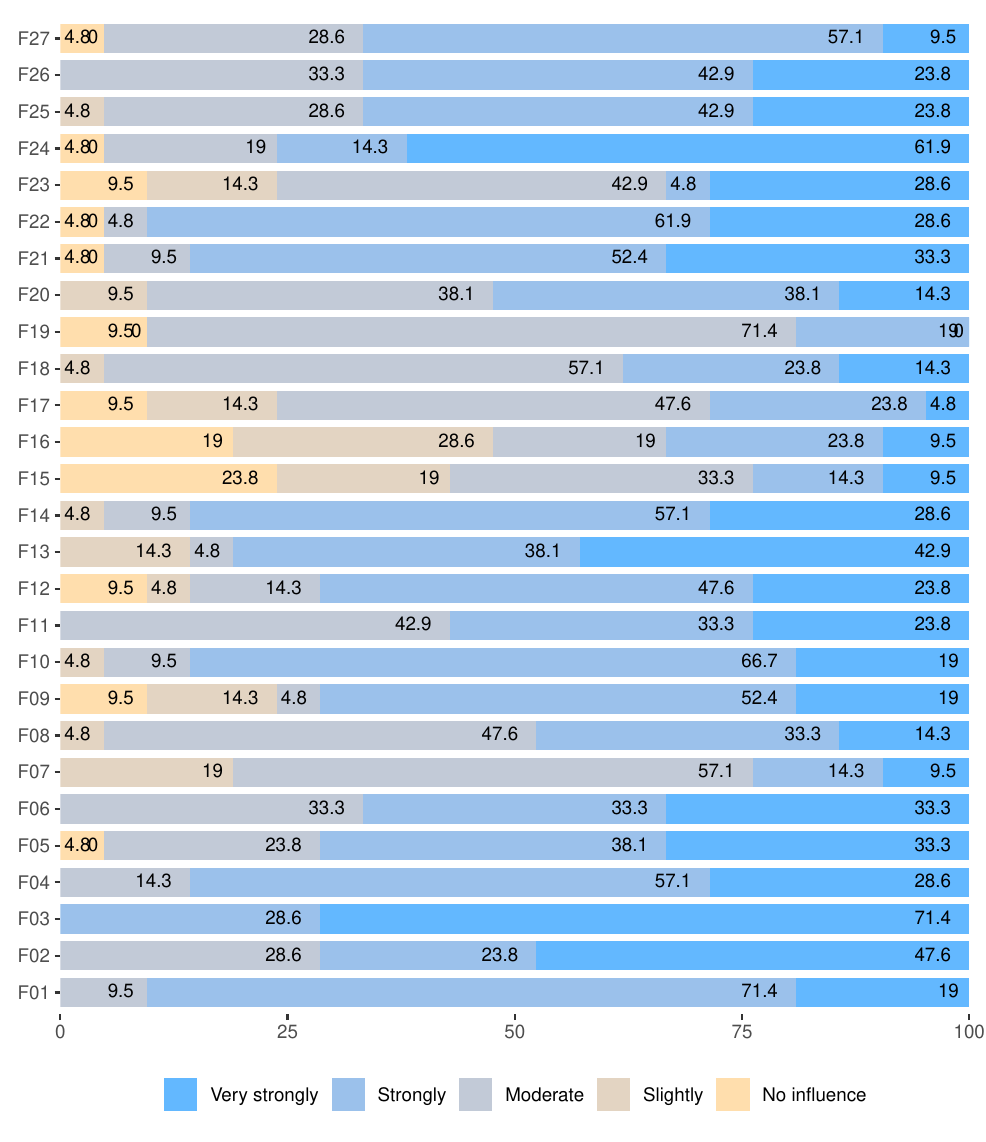}
    \caption{Frequency analysis for the factors under analysis and their influence on adopting or continuing to contribute to a software ecosystem (Round 2).}
    \label{fig:answers-round-2}
\end{figure}

We received 21 responses in Round 2. Figure~\ref{fig:answers-round-2} presents the new distribution of the selected factors' influence on a subject choosing a given ecosystem to start developing software. Each factor is represented in a row. Each row has three to five blocks that must be read from left to right to increase the influence level for the related factor. Numbers in the bars indicate the percentile of respondents who answered the question associated with the specific factor for the related influence level.

The set of 21 participants changed 111 answers compared to Round 1 out of 567 possibilities (19.5\%).The most changed answers were F12 (8 changes), F2, F3, and F13 (each with 6 changes), F1, F4 to F7, F19, F22, F24, and F26 (each with 5 changes). F9, F10, F16, F20, and F25 had their answers changed by 4 participants each, F5, F6, F8, F14, F15, F17, F18, and F27 had their answers changed by three participants and F11, F21, and F23 had their answers changed by two participants. 
 
Most (19 out of 21) participants changed their answers to at least one question. Four participants updated their answers to five questions, three changed their answers to three questions, and another three to seven questions. The cases with the most modifications were three respondents who changed their answers to 10, 12, and 14 questions, respectively (in the latter case, more than half of the questions).

Participants tended to increase the importance of factors in selecting an ecosystem (83 changes) in Round 2. The influence of the factors was changed from moderate to strong in 35 answers, from strong to very strong in 23 answers, from slightly to moderate in 15 answers, from moderate to very strong in 4 answers, from slightly to strong in 3 answers, from no influence to moderate in two answers and from slightly to very strong in one answer. Participants also tended to change their original answers by just one level (73 changes), with a shift in three levels observed in just one case (from slightly to very strong). This moderation in changing their responses can also be observed when the influence of the factors is reduced (28 responses). Fourteen changed their answer from strongly to moderate, 12 from very strongly to strongly, one from moderate to slightly, and one from slightly to no influence.

All the changes to factors F6, F9, F12, F14, F21, F22, and F27 aimed at increasing their influence level. 80\% of the changes to factors F1, F4, F24, and F26 had also the same intention. Something similar happened for factors F10, F16, F20, and F25 (75\% of the changes aimed to increase the level of influence of these factors),  F2, F3, F5, F8, F13, and F18 (66.7\%), F19 and F7 (60\%). Opinions regarding factors F11 and F23 were mixed: half of the participants opted to increase the degree of influence of these factors, while the other half reduced it. The only factor whose influence was reduced was F17 (33\% of the changes represented an increase in influence).

Round 2 of the analysis supports the results found in Round 1 in two aspects. Firstly, in Round 1, we found 18 factors in which at least half of the participants' answers indicated the importance level as ``strongly'' or ``very strongly''. We observed 19 factors with this characteristic in Round 2. The novelty was factor F20, which reached the limit after three changes increased its influence in Round 2. Secondly, if we only consider the factors with agreement above 75\% of responses as ``strongly'' or ``very strongly'', the six factors found in Round 1 (F1, F3, F10, F14, F21, and F24) remain on the list for Round 2. None of these factors was reduced in importance in Round 2; on the contrary, their importance increased in comparison with the original round. They were also joined by factors F4, F13, and F22. Therefore, we conclude that the second round of the Delphi study adds evidence to Round 1’s results.

To conclude, we turn back to research question SQ5: ``\textit{How do third-party developers evaluate the influence of DX factors to adopt and keep contributing to a SECO?}''. The Delphi study has shown that the same factors contribute to attracting new developers to a platform and keeping current developers actively contributing. Table~\ref{tab:ranking} presents a ranking (from most influential to least influential) of the DX factors by category, considering the indicator of agreement with the degree of strong influence ($SIP$). The most relevant DX factors to foster developer participation are \textbf{(F1) Desired technical resources for development}, \textbf{(F3) Financial costs for using the platform}, \textbf{(F4) Diversity of services provided by the platform}, \textbf{(F10) Application distribution methods}, \textbf{(F13) Ease of learning about technology}, \textbf{(F14) Low barriers to entry into the applications market}, \textbf{(F21) Emergence of new market and job opportunities}, \textbf{(F22) More financial gains}, and \textbf{(F24) Improvement of developers skills and intellect}. Thus, developers look forward to platforms with low entry costs, a wide range of services, comprehensive learning materials, and opportunities for financial gains. 

\begin{table}[!ht]
\centering\scriptsize
\caption{Ranking of ecosystem adoption and continued contribution influence indicator values for each factor by category.}
\label{tab:ranking}
\begin{tabular}{|p{5.2cm}|p{6.5cm}|c|}
\hline
\textbf{Category}  & \textbf{DX Factor} & \textbf{SIP} \\\hline

\multirow{8}{*}{Common Technological Platform} &   (F3) Financial costs for using the platform
        &  100.0   \\\cline{2-3}
 &  (F1) Desired technical resources for development
         &  90.4   \\\cline{2-3}
& (F4) Diversity of services provided by the platform
          &  85.7   \\\cline{2-3}
&  (F2) Easy to configure platform
         & 71.4    \\\cline{2-3}
&  (F5) Platform transparency
         & 71.4    \\\cline{2-3}
&  (F6) Documentation quality
         & 66.6    \\\cline{2-3}
&  (F8) Platform openness level
         & 47.6    \\\cline{2-3}
&  (F7) Existence of communication channels
         & 23.8    \\\hline                        

\multirow{6}{*}{Projects and Applications}

& (F14) Low barriers to entry into the applications market
         & 85.7    \\\cline{2-3}

&  (F10) Application distribution methods
         &  85.7   \\\cline{2-3}
&  (F13) Ease of learning about technology
         & 81.0   \\\cline{2-3}
 &  (F12) Requirements for developing applications over a platform
         & 71.4    \\\cline{2-3}

&         (F9) More clients/users for applications
        &  71.4  \\\cline{2-3}
 
& (F11) Application interface and appearance standards
          &  57.1   \\\hline

 \multirow{6}{*}{Community Interaction}                                                         &   (F20) Community size and scalability
        & 52.4    \\\cline{2-3}
             & (F18) Knowledge exchange between community developers
          &   38.1  \\\cline{2-3}
         &  (F16) Commitment to the community
         &  33.3   \\\cline{2-3}
          &    (F15) (F15) Obtaining community recognition
       &  23.8   \\\cline{2-3}
               &   (F19) A good developer relations program
        & 19.0    \\ \hline  

\multirow{7}{*}{Expectations and Value of Contribution}                   
 &   (F22) More financial gains
        &  90.5   \\\cline{2-3}
&    (F21) Emergence of new market and job opportunities
       &  85.7   \\\cline{2-3}
&  (F24) Improvement of developer skills and intellect
         &   76.2  \\\cline{2-3}
& (F26) Qualities and characteristics of the SECO platform
          &   66.7  \\\cline{2-3}
 & (F25) Autonomy and self-control of workflow
          &   66.7  \\\cline{2-3}
   &  (F27) Engagement and rewards for work
         &  66.6    \\\cline{2-3}    
  & (F23) Fun while developing
          &   33.4  \\\hline
\end{tabular}
\end{table}

\section{Discussion and Implications}\label{sec:discussion}

Some studies conducted in academia and industry related to DX characteristics can vary greatly. As mentioned by Nylund~\cite{Nylund2020}, research on this concept has increased in recent years, a fact that suggests DX is an emerging term. When we consider the scope of SECO, few studies are focusing on the investigation of DX, although this concept is crucial for many activities within a SECO.

During the selection of studies in our SMS, we observed that many of them had a definition for DX based on seniority level. In other words, how much a developer can be considered experienced in performing a task. This plurality of meanings related to the term DX was intriguing when trying to understand research on the topic. This work chose to consider the definition of DX by Fagerholm and Münch \cite{Fagerholm2012} in the selection process, excluding studies that did not fit in this view.

Considering that the work by Fagerholm and Münch \cite{Fagerholm2012} was published in 2012, we observe that the 29 selected studies are comprised between the years 2013 to 2023. We believe this can be explained because studies focusing on SECO become more common over the last decade. Corroborating with Nylund~\cite{Nylund2020} and Steglich et al.~\cite{Steglich2019}, the analysis, interpretation, and compilation of DX factors is somewhat complex, as many studies implicitly address the term, mainly in older works. Therefore, the research strategy adopted in this work sought to bring the maximum number of studies that could be filtered for the elaboration of Figure~\ref{fig:ListDxFactors}.

Regarding the research methods used by the selected studies to identify the DX factors, we observed that the majority opted for conducting interviews and questionnaires. Interviews allow for greater depth in the analyses, on the other hand, questionnaires allow for reaching a large number of participants. Given the intrinsic subjectivity of research involving DX, these methods are successful because they enable collecting the perception of those who experience it daily.

The importance of DX in SECO has gained more strength as the need to attract and keep developers actively contributing to a platform has become essential for SECO sustainability (S5, S8, S9, S10, S21, S22, and S24). Based on this need, we conducted the Delphi study to understand how DX factors impact them in their daily routines, mainly when choosing a SECO and keeping contributing. These are one of the main concerns for keystones to increase and maintain indicators of SECO health and sustainability (S2, S3, S6, S7, S12, S13, S14, S17, S18, S19, S21, S23, and S25). By analyzing them from the perspective of software developers who work within SECO, we can view them more pragmatically, in addition to confirming the findings of the literature.

Based on the responses of the 21 participants, the Delphi study allowed them enough time to think about each one of 27 DX factors identified in the literature, without making participation in the study exhaustive. We realized that the dynamics of two evaluation rounds allowed participants to reconsider their choices before giving a definitive answer, which was a positive aspect so that we could characterize the influence of DX factors. Next, we discuss how DX factors can be addressed in the concerns presented in SQ4 in Section~\ref{sec:resultsSMS} and give some recommendations for keystones.

\vspace{8px}
\noindent
\textbf{Attracting or maintaining developers in SECO over time}: we recommend that keystones pay attention to factors that, according to participants, have the most influence on attracting and retaining developers at SECO. To achieve this, keystones should:

\begin{itemize}
    \item Provide desired technical resources for development (F1): invest in robust infrastructure and tools that meet the needs of third-party developers, such as API, libraries, and IDE. For example, AWS provides a comprehensive set of tools that facilitate development at various stages, from build to deployment;
    \item Promote fair financial costs for using the platform (F3): offer scalable and transparent pricing models, with free or low-cost options for beginners and affordable pricing for advanced features. This approach will make the platform accessible to a wider range of third-party developers, similar to how GitHub offers free plans for open-source projects;
    \item Decrease barriers to entry into the applications market (F14): simplify the sign-up and publishing processes, ensuring that developers can get started quickly with minimal friction. This includes providing clear documentation and tutorials, similarly to Google Play Store, which offers an accessible and straightforward publishing process for third-party developers;
    \item Allow third-party developers to obtain more financial gains (F22): create clear and fair monetization opportunities that allow developers to monetize their creations. A model similar to the Unity Asset Store\footnote{https://assetstore.unity.com/}, where developers can sell assets and receive a significant share of the profits, could be beneficial.
\end{itemize}

\noindent
\textbf{Ensuring SECO sustainability}: in addition to considering the factors mentioned in the previous concerns, the keystone must create conditions for productivity to increase it in SECO. Therefore, keystones should:

\begin{itemize}
    \item Provide a diversity of services in a SECO platform (F4): providing a wide range of services that meet different needs within a SECO, such as scalable infrastructure, developer tools, and support services. This will increase the productivity and satisfaction of third-party developers, making a platform more attractive for long-term use and growth. For example, AWS and Google Cloud\footnote{https://cloud.google.com/} provide flexible infrastructure, allowing developers to scale their applications based on usage and need;
    \item Provide an efficient application distribution (F10): ensuring a transparent, accessible, and effective process for distributing applications within SECO. This includes creating a marketplace for easy application deployment, providing mechanisms for updates, and providing visibility for high-quality applications. Efficient distribution helps increase the reach of developers’ work, improving the overall health of the ecosystem. For example, the Google Play Store and Apple App Store provide a platform where developers can distribute their applications to a global audience, helping users to discover and install applications.
\end{itemize}

\noindent
\textbf{Defining the level of SECO openness}: this recommendation concerns several approaches to ensure that a SECO platform is accessible, transparent, and attractive to third-party developers. To do so, keystones should:

\begin{itemize}
    \item Invest in strategies to promote adequate transparency on a SECO platform (F5): promoting transparency by considering open source practices, such as those used by GitHub and Linux\footnote{https://www.linux.org/}, and making roadmaps and platform updates publicly available. This will foster trust and engagement in the developer community;
    \item Improve the documentation quality (F6): investing in clear, detailed documentation with interactive tutorials and code examples will ensure that third-party developers can easily understand and integrate applications to the SECO platform;
    \item Invest in the process of learning SECO platform technology (F13): providing educational resources such as courses and tutorials and fostering active developer communities such as those seen with React\footnote{https://react.dev/} will help new developers quickly adapt themselves to the platform. Finally, implementing efficient support channels and fostering frequent feedback loops will ensure that the platform evolves in response to user needs, fostering a sustainable ecosystem.  
\end{itemize}

\noindent
\textbf{Dealing with power relations within SECO}: although the factors in the community interaction category did not show high influence indicators, we recommend that keystones should:

\begin{itemize}
    \item Promote developer relations programs (F19): keystones should promote open communication and transparency, similar to Google Developer Relations or the GitHub Sponsors program\footnote{https://github.com/sponsors}, in which developers are not only supported but also encouraged to contribute to the evolution of the SECO platform. Furthermore, providing clear and accessible documentation and resources will help mitigate power imbalances by equipping all developers, regardless of their experience level, with the tools they need to engage meaningfully;
    \item Investing in community proximity and scalability (F20): engaging with the community is important to understand third-party developers’ needs and address power imbalances within the ecosystem. Keystones should actively promote inclusive feedback mechanisms and decision-making processes to ensure that diverse voices in the community are heard. This can be achieved through open forums, developer surveys, and community-led initiatives. By ensuring that all developers feel represented and empowered, keystones can mitigate power dynamics that may marginalize certain groups and foster community growth.
\end{itemize}

\section{Threats to Validity}\label{sec:threats}
This work presents two different studies (SMS and Delphi) and each of them has specific threats and limitations. Thus, we report the identified threats and strategies to mitigate them.

Some threats to the validity of this SMS were identified. Our analysis considers descriptive validity, theoretical validity, generalizability, and interpretive validity, according to Petersen et al.~\cite{Petersen2015}. During the course of this research, we sought to minimize the influence of these threats and reduce their possible risks. \textit{Descriptive validity}: to reduce this threat, a data collection form has been designed to support the recording of data to answer the questions. \textit{Theoretical validity}: since there is no consensus in SE and SECO on the definition of DX, we chose to select and analyze the studies under the aegis of DX frameworks by Fagerholm and Münch~\cite{Fagerholm2012} and Greiler et al.~\cite{Greiler2022} to minimize possible interpretative issues. The search string was defined inclusively to capture studies related to DX in SECO, but we also recognize that we could have applied the forward snowballing and included more sources to try to increase the number of selected studies.

\textit{Generalizability}: generalization is not a huge threat once we have used a structured protocol based on Petersen et al.~\cite{Petersen2015}, which facilitates replication. We also make available the datasets in the supplementary material. \textit{Interpretive validity}: to minimize the researchers’ bias, when there was doubt in executing the selection process, this was discussed between two researchers extensively and the differences were analyzed together with a third researcher until there was a consensus. It is worth highlighting that the protocol for SMS does not consider the quality of the retrieved studies.

Concerning the Delphi study, our analysis considers construct validity, internal validity, external validity, and reliability, according to  Wohlin et al.~\cite{Wohlin2012}. \textit{Construct Validity}: to mitigate threats related to construct validity in the DX factors evaluation, such as the inadequate design of the questionnaire, bias in expert
selection, and overreliance on participants' opinions, we performed a Delphi study. with the following strategy: (i) we carried out a pilot to tune the questionnaire and reduce the ambiguities in the questions; (ii) we anonymized responses to reduce bias; and (iii) we supplemented participants' opinions with empirical data whenever possible.

\textit{Internal Validity}: refers to uncontrolled factors that may affect the study results~\cite{Wohlin2012}. It consists of the results’ equivalence with the reality or the degree to which the study minimizes bias. To mitigate these threats, we previously defined a research protocol with objective criteria related to the selection of participants in our Delphi study and we carried out a pilot to ensure the clarity of the questionnaire.

\textit{External validity}: refers to the capacity to generalize the results to other contexts, populations, or settings beyond those directly examined in the study~\cite{Wohlin2012}. It examines whether the results observed in a specific context are applicable in other situations or among different groups of people. To mitigate these threats, we non-probabilistic convenience sampling technique, following the guidelines of Kitchenham et al.~\cite{Kitchenham2015}, to select participants who adequately meet the profile defined for the study. The strategies for attracting participants were specifically aimed at third-party developers who interact with any SECO to develop their applications, in addition to the snowball sampling technique, in which the first participants nominated other professionals to participate. Although the Delhi study is more quantitative, the iterative response process allows inappropriate responses to be revised, allowing for a more qualitative and interpretive analysis of the results.

\textit{Reliability:} refers to the consistency and repeatability of measurements or observations obtained in a study~\cite{Wohlin2012}. This aspect addresses how much the data and analysis rely on the specific researchers involved. To reduce potential biases, we carried out multiple rounds of review and discussion among this study's authors regarding the definition of the DX factors. Conflicts on data analysis results were resolved through collaborative discussions until a consensus was achieved.

\section{Conclusion}\label{sec:finalRemarks}


This work aimed to identify key DX factors and understand how they influence third-party developers' decisions to adopt and keep contributing to a SECO. To do so, we conducted an SMS and selected 29 studies to analyze the state-of-the-art of DX in SECO. Next, we conducted a Delphi study to evaluate the influence of 27 DX factors (identified in our SMS) from the perspective of 21 third-party developers to adopt and keep contributing to a SECO. As the main contribution, we provided a set of 27 DX factors organized into four categories that affect DX in SECO to help researchers and practitioners have a better understanding of the topic. 

Regarding implications, researchers can find in this work an overview of what has been studied so far about DX related to the SECO context. The categorization of factors can function as a starting point for future research on more specific factors, such as exploring only factors related to development infrastructure or ecosystems. By identifying the factors, practitioners can improve DX by knowing what affects their daily lives and understanding the effects in SECO. We also provide recommendations for keystones and practitioners about how to use the list of DX factors as a consultation guide to deal with the main concerns related to DX in SECO.

The main takeaway message of this paper is that research on DX in SECO is very promising but can still be expanded. There are few studies focused on the topic, but the concept of DX is crucial to the dynamics of SECO, promoting its health and sustainability. Our set of DX factors can be used as a reference guide and recommendations to support practitioners from keystones dealing with the key concerns related to DX in SECO.

For future work, we suggest conducting case studies to explore the factors of each of the categories presented in this work. It is still necessary to understand and map the benefits of each factor for both developers and keystone, with the possibility of creating a catalog of actions to improve DX in SECO.

\section*{Acknowledgements}
This work was financed in part by the Coordenação de Aperfeiçoamento de Pessoal de Nível Superior – Brazil (CAPES) – Finance Code 001 and Grant 88887.928989/2023-00, CNPq (Grant 316510/2023-8), FAPERJ (Grant E-26/204.404/2024), and UNIRIO.

\bibliographystyle{elsarticle-num} 
\bibliography{bibliography}

\end{document}